# Room temperature current suppression on multilayer edge molecular spintronics device

Pawan Tyagi,

Department of Chemical and Materials Engineering, University of Kentucky, Lexington, Kentucky-40506, USA
Current Address: School of Engineering and Applied Science, University of the District of Columbia, Washington DC-20008, USA

**Abstract:** Molecular conduction channels between two ferromagnetic electrodes can produce strong exchange coupling and dramatic effect on the spin transport, thus enabling the realization of novel logic and memory devices. However, fabrication of molecular spintronics devices is extremely challenging and inhibits the insightful experimental studies. Recently, we produced Multilayer Edge Molecular Spintronics Devices (MEMSDs) by bridging the organometallic molecular clusters (OMCs) across a ~2 nm thick insulator of a magnetic tunnel junction (MTJ), along its exposed side edges. These MEMSDs exhibited unprecedented increase in exchange coupling between ferromagnetic films and dramatic changes in the spin transport. This paper focuses on the dramatic current suppression phenomenon exhibited by MEMSDs at room temperature. In the event of current suppression, the effective MEMESDs' current reduced by as much as six orders in magnitude as compared to the leakage current level of a MTJ test bed. In the suppressed current state, MEMSD's transport could be affected by the temperature, light radiation, and magnetic field. In the suppressed current state MEMSD also showed photovoltaic effect. This study motivates the investigation of MEMSDs involving other combinations of MTJs and promising magnetic molecules like single molecular magnets and porphyrin. Observation of current suppression on similar systems will unequivocally establish the utility of MEMSD approach.

**Introduction:** Molecular spintronics devices (MSDs) can revolutionize the computer's logic and memory [1]. A MSD is a highly promising platform to enable the quantum computation [2, 3]. The MSD's functioning depends upon manipulation of the spin degree of freedom of electron(s), requiring small energy for their manipulation. Such spin devices are expected to work with significantly lowered energy input, as compared to the charge-based devices. Since a molecule is the functional unit of a MSD, hence devices can be miniaturized up to the molecule's length scale. Moreover, MSD design enables the utilization of a large variety of molecules [4-6] as the device element between the two ferromagnetic (FM) electrodes [7, 8]. The simplest form of a MSD is analogous to magnetic tunnel junction (MTJ) based spin valves. Petta et al.[7] developed MTJ using octanethiol molecular tunnel barriers and nickel (Ni) FM electrodes. These molecule based MTJs showed <10% magneto resistance (MR) when the direction of magnetization of the two FM electrodes were changed from parallel to anti-parallel; the transport measurements were conducted at ~4.2 K. Interestingly, theoretical calculations have predicted several hundred percent MR ratio with the molecule based MTJ [1].

Novel MSDs are expected to evolve from a system of FM electrodes and molecule with a net spin. A number of intriguing phenomenon have been theoretically predicted for such systems [9-11]. For instance, Petrov et al. [10] predicted ~7 orders resistance change for the molecular bridges, which were initially antiferromagnetically coupled with the electrodes. The single molecular magnets and quantum dots with a net spin state were indicated to be promising for the experimental realization of this radical predication [10]. Laurenberg et al. [11] made another interesting theoretical prediction about the influence of the interference of molecule's quantum tunneling paths on MSD's transport. The single molecular magnets connected between the two FM electrodes were predicted to exhibit Berry phase interference [11]. Berry phase interference was shown to influence the evolution and the quenching of Kondo resonance. *Molecules are akin to quantum dots*. Martinek et al. [12, 13] predicted the Kondo resonance for a system enabling the interaction between the spins of FM electrodes and the spin of quantum dots. This observation was experimentally realized when $C_{60}$ molecule(s) bridged the nanogap between the two FM electrodes, on a Ni break junction [14]. Interestingly, Ni break junction also showed the splitting of Kondo resonance in zero external magnetic field [14]. Strong magnetic coupling between the spin states of $C_{60}$ molecule and the FM electrodes produced >50 T local magnetic field to yield the Kondo level splitting. Utilization of numerous theoretical insights has not been possible due to the negligible progress in the fabrication of MSDs.

Due to fabrication difficulties in producing reliable FM contacts to the magnetic molecules and failure in performing extensive control experiments the field of MSD is still in embryonic stage [8]. Recently, we produced the multilayer edge molecular electronics devices (MEMEDs) [15, 16]. For the fabrication of a MEMED, the molecular conduction channels were bridged across the insulator of a prefabricated -tunnel junction, along its edges. If tunnel junction is a MTJ, which have two FM electrodes, then MEMED will become a multilayer edge molecular spintronics device (MEMSD) [16] (Fig. 1a). A MEMSD can be formed by using a vast permutations of molecules and FM electrodes. A MEMSD approach enables an unprecedented number of control experiments, such as reversing the molecule effect and characterizing a MTJ before transforming it into a molecular device [16]. The MTJ used for MEMSD can be subjected to magnetic characterizations before and after treating it with desired molecules. Such magnetic characterizations can be correlated with the transport properties of a MEMSD. Magnetic and transport studies together can be crucial in bringing the complete understanding of MEMSDs. This paper discusses the fabrication and characterization of the MEMSDs, which showed as much as six orders current suppression at room temperature. Organometallic molecular clusters (OMCs) across the AlOx insulator of MTJ with cobalt (Co)/80% nickel-and 20% iron alloy (NiFe)/ alumina (AlOx)/NiFe configuration (Fig. 1a) exhibited dramatic changes in MTJ's magnetic and spin transport properties. This paper focuses on the transport properties and discusses the supporting experiments to establish the observation of current suppression.

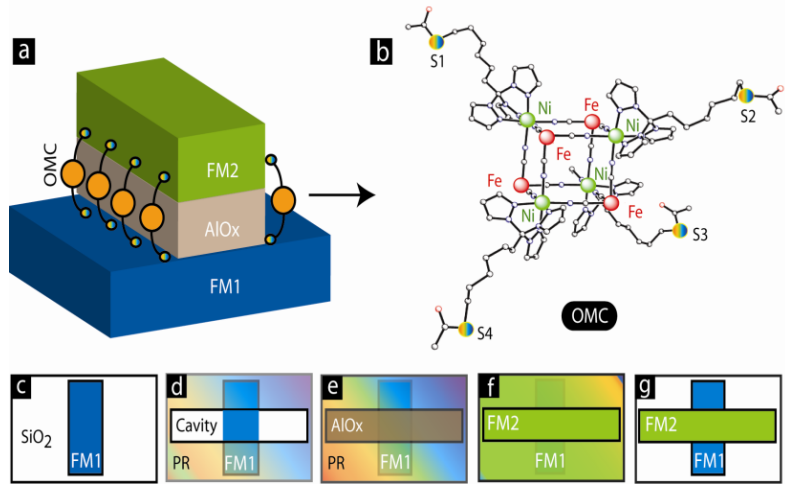

Fig.1: (a) A MEMSD is produced by bridging (b) OMCs across the insulator of a prefabricated MTJ. Fabrication steps for MEMSD are following: (c) deposit the first FM electrode (FM1) on insulating substrate, (d) create photoresist (PR) cavity pattern for the deposition of (e) ~2 nm AlOx and (f) top FM electrode (FM2). (g) liftoff step produces a MTJ with the exposed sides where OMCs are bridged across the insulator to produce (a) MEMSD.

**Experimental details:** The MTJs for MEMSDs were fabricated on thermally oxidized silicon (Si). The bottom FM electrode (FM1) was generally a bilayer comprising 5-7 nm thick Co and a 3-5 nm thick NiFe (Fig. 1c). Next, photolithography was performed to create a cavity (Fig. 1d) for the deposition of a 2 nm thick AlOx (Fig. 1e) and a ~10 nm thick NiFe top electrode (FM2) (Fig. 1f), respectively. The deposition of AlOx and FM2 via the same photoresist (PR) cavity ensured that along the MTJ edges the minimum gap between the two FM electrodes is equal to the AlOx insulator thickness (Fig. 1a and g). The liftoff of PR produced Co/NiFe/AlOx/NiFe MTJ with the exposed side edges (Fig. 1g). Along the exposed side edges OMCs (Fig. 1b)[4] were bridged across the AlOx to complete the MEMSD fabrication. These OMCs exhibited S=6 spin state in the bulk powder form at <10 K. A OMC possessed cyanide-bridged octametallic molecular cluster, $[(pzTp)Fe^{III}(CN)_3]_4[Ni^{II}(L)]_4[O_3SCF_3]_4$ [(pzTp) = tetra(pyrazol-1-yl)borate; L = 1-S(acetyl)tris(pyrazolyl)decane][4] chemical structure. With the help of thiol functional groups, an array of OMCs was covalently-linked onto the NiFe layer of the top and bottom electrodes. Specifically, OMC channels were electrochemically bridged on to the FM1 and FM2 surfaces [17]. For the molecule attachment, MTJ samples were immersed in a dichloromethane solution of OMCs (0.1 mM). An alternating ±100 mV bias with a time interval of 0.002 seconds for 2 min was applied between the two FM electrodes [15]. After electrochemistry step, resultant MEMSDs were rinsed with dichloromethane, 2-propanol, and DI water, respectively. Lastly, samples were dried under a nitrogen gas stream. MEMSD device fabrication details are also published elsewhere [15, 18].

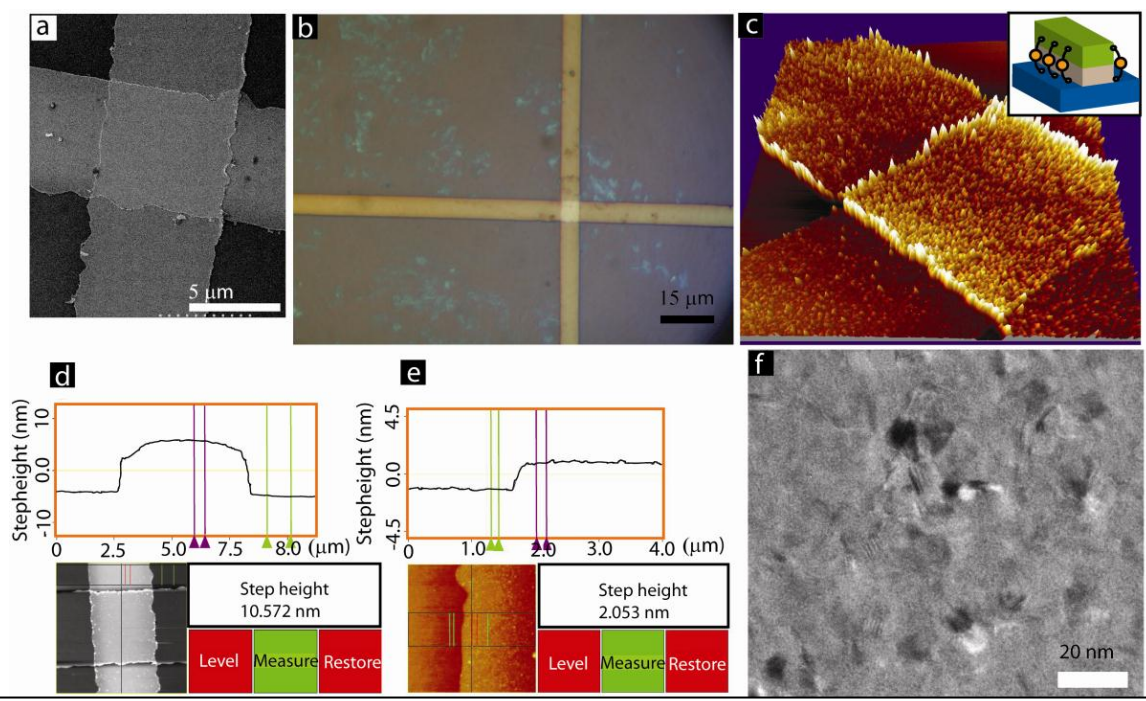

Fig. 2: Physical characterizations: (a) SEM of a typical MEMSD, (b) optical micrograph of a Ta/Co/NiFe/AlOx/NiFe tunnel junction after the electrochemical step for the OMC bridging, (c) 3D AFM image showing side edges of a tunnel junction along which OMCs are bridged across insulator, in the manner shown in the inset of (c). (d) AFM based thickness and edge profile measurement of bottom Co(5nm)/NiFe(5 nm) electrode. (e) AFM based thickness measurement of AlOx insulator. (f) TEM study showing the representative topology of AlOx utilized in MEMSD.

NiFe FM film was extensively employed in MEMSDs. NiFe electrode possessed several useful attributes to enable the fabrication of a MEMSD: (i) NiFe is ambient stable and start oxidizing upon heating around 200 ºC [19], (ii) thermodynamically, on NiFe surface only iron atoms are oxidized and Ni atoms remains in the elemental state [19]; Ni atoms can covalently bond with the molecules [15]. (iii) NiFe is unaffected by the molecular solution, and electrochemical protocol used for molecule attachment. NiFe serves as an excellent protection for other etching susceptible metals like Co. (iv) NiFe deposited on the other FM films like Co produced different magnetic properties than that of NiFe alone; the magnetization for a bilayer Co/NiFe and NiFe saturated at ~60 Oe and ~15 Oe magnetic field, respectively. The role of Co was mainly to produce a Co/NiFe magnetic electrode with different magnetic properties, as compared to NiFe. The thickness of Co was typically kept in 5-7 nm range; Co with > 10 nm thickness made MEMSD unstable. A Co metal, with > 10 nm thickness, produced nanohillocks and punctured film(s) right above it [20]. These nano hillocks not only created the short circuit between the two electrodes but also made Co/NiFe electrode prone to localized etching, at the sites of naohillocks. Besides, Co/NiFe/AlOx/NiFe we also studied MEMSD with NiFe(10 nm)/AlOx(2 nm)/NiFe(10 nm) and Ta (5 nm)/Co(5 nm)/NiFe(5 nm)/AlOx (2 nm)/NiFe (10 nm)/Ta(5 nm) MTJ configurations.

Extensive physical characterizations were performed to optimize a MEMSD (Fig. 2) with an aim of reducing leakage current through the planar area of the MTJ test bed. The bottom electrode edges were tapered (Fig. 2d) by using a slight undercut in photoresist pattern during the first photolithography step. Tapered edge profile ensured that notches do not appear on the edges of the bottom FM electrode to create short circuits between the two FM electrodes. Additionally, it was also found that the use of tantalum (Ta) seed layer helped producing a smoother morphology; a smoother morphology of bottom FM electrode (< 0.2 nm RMS roughness) enabled lower the leakage current through ~2 nm AlOx tunnel barrier. Furthermore, MTJs were annealed at 200 ºC for > 2 hours in nitrogen ambience to reduce the leakage current [21, 22]. It was also critical to minimize the notches along the side edges of AlOx to increase the chances of bridging molecules across the insulator gap. To optimize the AlOx edge profile, the photolithography and the sputtering deposition steps were customized. The quality of MTJ and AlOx edges were highly susceptible to the photoresist quality.

Transport studies of the MEMSD were performed with a Keitlhley 2430 1kW pulse source meter and Keitlhley 6430 sub-femtoamp source meter. Samples were mounted on a metallic chuck, located in a faraday cage. Biaxial and triaxial cables were used to electrically connect the probe needles to a source meter. As a standard procedure current-voltage (I-V) measurements were performed before and after molecule attachment. Generally, I-V studies were performed in ±100 mV bias range; use of a low bias range did not appear to induce instability, as caused by the high bias application. Typical AlOx tunnel barrier breakdown voltage and cross section area was ~1.7 V and 10-50 $\mu m^2$, respectively (Fig. 2 a-b).

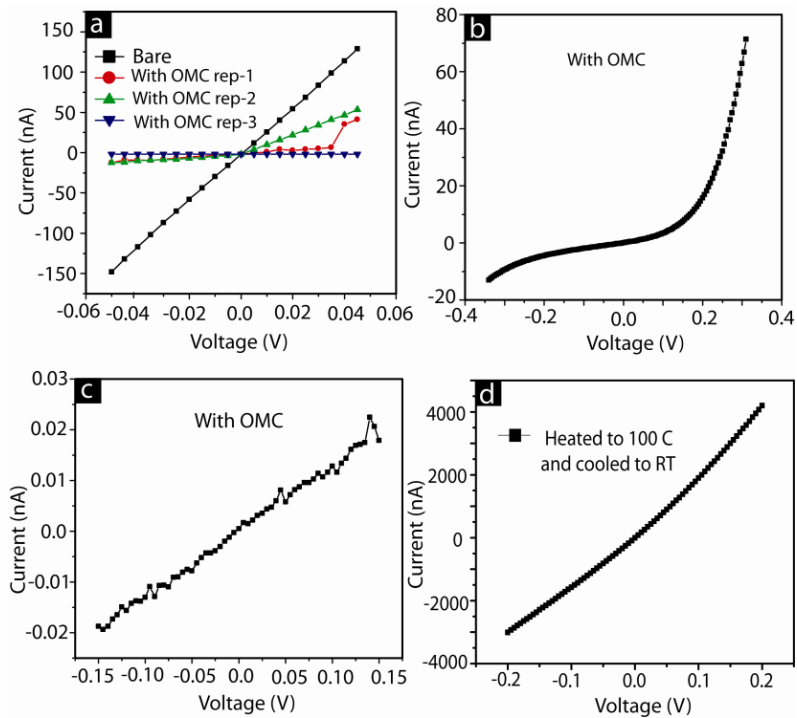

Fig. 3: Different stages of current suppression on the first MEMSD: (a) repeating (rep-1,2,3 etc.) I-V three times in transient low current state induced by OMCs, (b) stabilized OMCs induced asymmetric I-V, (c) effect of incubation further lowered the current response, (d) heating the sample to 100 ⁰C and then cooling it back to RT produced the high current state.

**Results and discussion:** The I-V response of several MTJs with Co/NiFe/AlOx/NiFe configuration showed the OMC induced current suppression. This paper discusses several cases to establish this observation. In general, right after OMCs attachment along the MTJ edges a transient current lowering was observed for several hours. Eventually, a suppressed current state stabilized. The first MEMSD discussed in figure 3 exhibited multiple OMC induced current states. The first MEMSD utilized a MTJ with [Ta(5 nm)/Co(5 nm)/NiFe(5 nm)/AlOx(2

nm)/ NiFe (10 nm) ] configuration and exhibited a labile current lowering after the bridging of OMCs (Fig. 3a). After several hours, I-V profile became asymmetric and settled in the nA range (Fig.3b). Leaving this MEMSD sample intact for 2-3 days brought the current down to ~10 pA range (Fig. 3c). At this point, it can be suspected that a mechanical damage in the electrical leads to the junction can produce such a current suppression. However, the same MEMSD regained a high current (μA) state upon heating to 100 ºC and then cooling back to room temperature (Fig. 3d). If serendipitous mechanical damage caused the current suppression, then it was unlikely that heating could weld the breakage at high temperature and maintain the electrical connection to the room temperature. Importantly, after few hours the temperature induced high current state *again* changed to a low current state. In addition to temperature, the magnetic field also influenced the suppressed current state. The magnetic field dependent transport on MEMSDs is discussed elsewhere in this study.

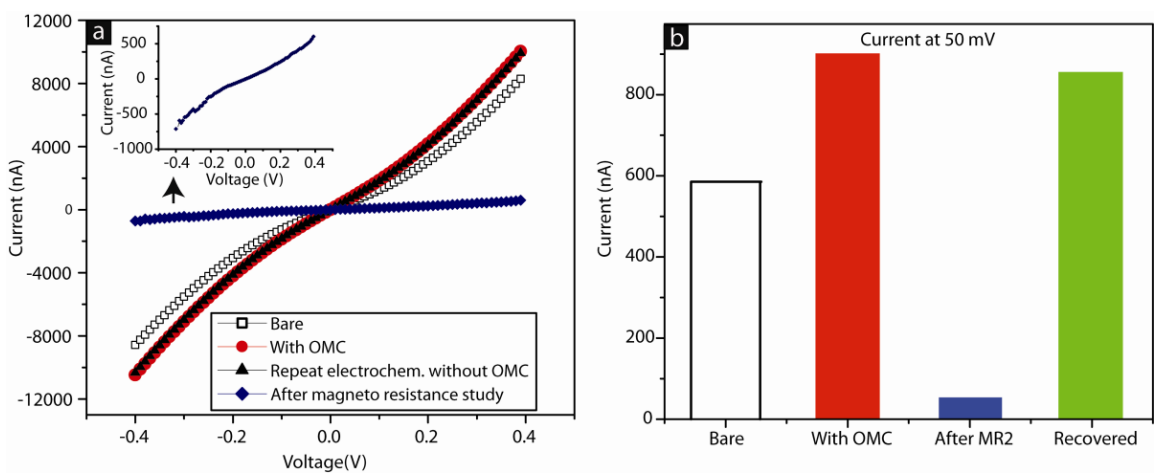

Fig. 4: The second MEMSD showing current suppression during magneto resistance (MR) study: (a) OMCs produced a stable current increase over the MTJ leakage level. (b) Net current at 50 mV in various stages.

If the results in figure 3 are free from all kind of artifacts, then it is a highly intriguing to see how ~10,000 OMCs along the MTJ's edges shut down the leakage current through the planar area. In fact, several nonmagnetic tunnel junction showed current increase due to the OMCs channels [15]. *A plausible reason behind the current suppression may be due to the OMCs induced changes in the basic magnetic properties of the FM electrodes; presumably due to the enhanced exchange coupling between two FM electrodes*. In an analogous system with ferromagnet-non magnet-ferromagnet enhancement of magnetic coupling affected the Curie temperature of the involved FM electrodes [23]. Here, if OMC are capable of dramatically increasing the exchange coupling between the FM electrodes of the Co/NiFe/AlOx/NiFe MTJ then equally dramatic transport characteristics are expected on the resultant MEMSD. We further discuss the underlying mechanism behind the current suppression elsewhere in this study.

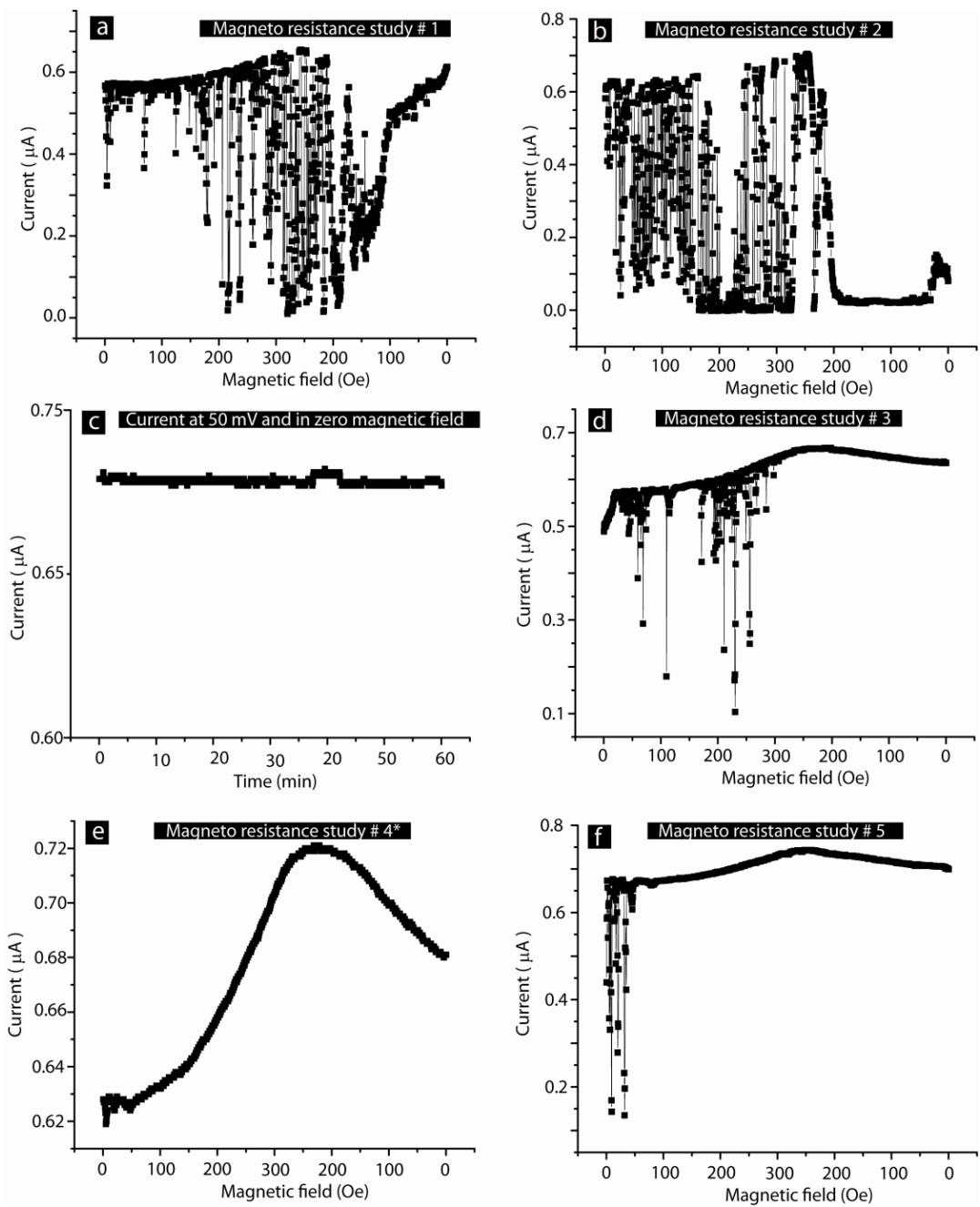

Fig. 5: Magneto resistance (MR) on the second MEMSD: During MR study the in plane magnetic field was increased from 0 to 300 Oe, and then reduced to 0 Oe. (a) The first MR showed occurrence of unstable low current states, (b) During the second MR a low current state stabilized. In the low current state, the repeating of I-V studies brought the MEMSD current back to µA range. (c) This µA current range remained stable when no magnetic field was applied. (d) The third MR study showed the occurrence of unstable low current states. (e) Rotating the sample by 90º yielded a clear MR curve. (f) The frequency of occurrence of low current state kept decreasing in subsequent MR studies.

A second MEMESD sample showed current suppression during the magneto resistance (MR) study. Here, current suppression did not occur automatically; it is noteworthy that the first MEMSD settled in the suppressed current state rather abruptly (Fig 3). In the present case, initially, OMC produced a stable current increase when OMCs bridged along the edges of a (Co (5 nm)/NiFe (5 nm)/AlOx (2 nm)/NiFe (10 nm) MTJ (Fig. 4a). To ensure that OMC channels, not the serendipitous defects arising from the OMC attachment protocol produced this change a number of control experiment were performed. Electrochemical protocol, without OMCs, was conducted on the second MEMSD sample. No change in the second MEMSD's current was observed (Fig. 4a). However, during the MR study this sample attained a low current state (Fig. 5). During MR study, this second MEMSD showed the occurrence of an unstable low current state (Fig. 5a). Interestingly, this sample gained a rather stable low current state during the second MR study (Fig. 5b). Immediately after the observation of this low current state, after the second MR study, I-V studies were conducted. I-V data in the inset of figure 4a correspond to the stabilized low current state. It appeared that in-plane magnetic field was assisting the stabilization of suppressed or low current state. This magnetic field induced low current state is in agreement with the effect of magnetic field observed on other MEMSDs. For instance, the first MEMSD sample exhibited 2-3 orders current suppression when it was subjected to the static in-plane 0.45 T magnetic field (Fig. 3). Interestingly, repeating I-V measurements disrupted the low current state and the second MEMSD returned into a high current state (Fig. 4b). The second MEMSD's current did not change noticeably at 50 mV in the absence of magnetic field (Fig. 5c). To confirm that a magnetic field play decisive role in bringing about a low current state MR studies were repeated. In the third MR study also dips of unstable low current state occurred (Fig. 5d). Interestingly, the MR study performed after rotating the MEMSD sample appeared without low current dips (Fig. 5e). However, repeating the MR study after restoring the sample position, as used in MR1-3, again produced the low current dips in the early stage of this study (Fig. 5f). In the subsequent MR studies the frequency of occurrence of the low current states gradually vanished. One can argue that such transient low current states (Fig. 5) could be due to the presence of atomic scale defects within the AlOx tunnel barrier. Undoubtedly, like OMCs the atomic defects can strongly influence the exchange coupling between the two FM electrodes [24, 25]. However, in our studies a defective tunnel barrier generally kept deteriorating to the state of a complete failure [20]. In the present case MEMSD maintained similar transport characteristics during multiple I-V and MR studies. To discuss the likely reason why the first MEMSD (Fig. 3) showed a stable current suppression but not the second MEMSD (Fig.4), we have following hypothesis. The main expected reason is the difference in the number of OMC channels on the two MEMSDs. In the latter case, we expected that the population of OMC channels along the MTJ's edges is insufficient to produce a stable current suppression. The other possible reason may be that the MTJ used for the second MEMSD was rather leaky, and hence did not allow OMC to show their full effect [15].

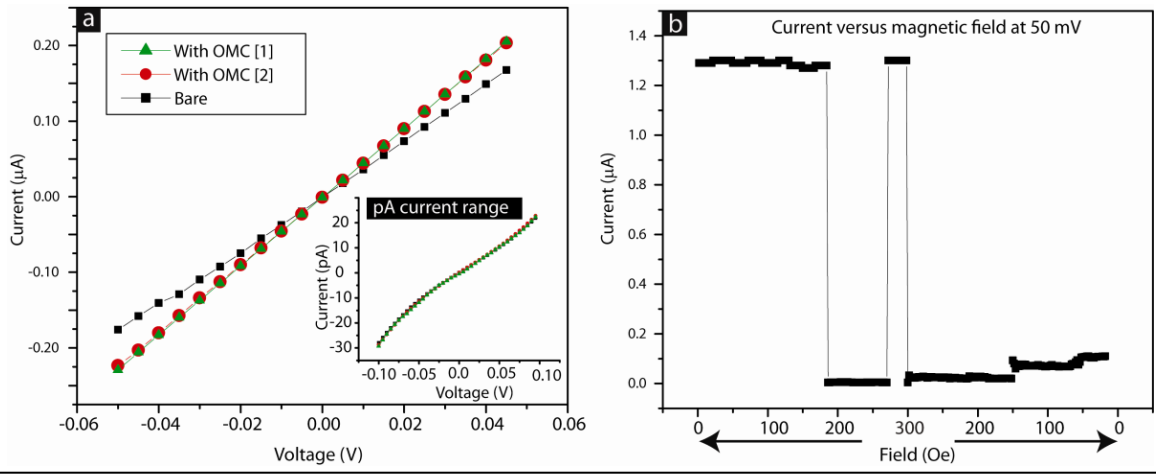

Fig. 6: Current suppression on the third MEMSD: (a) OMCs marginally increased the current of MTJ; after few experiments and magneto resistance study a suppressed current state stabilized (inset of panel a). (b) Magneto resistance study on this junction showed a clear switching of the current state between nA and μA level.

On the third MEMSD sample a clear current suppression and the cycling between the suppressed and the high current state was observed. Initially, OMCs increased the Co/NiFe/AlOx/AlOx/NiFe MTJ current (Fig. 6a). After several hours of OMCs attachment current remained unstable. During one MR study this MEMSD tend to settle in a several order smaller current range, as compared to the starting μA current range (Fig. 6b). Finally, this third MEMSD assumed a suppressed current state in pA range. It is noteworthy that, the MEMSD samples discussed in figure 5 and figure 6 has two similarities: (a) OMCs first increased the MTJ current and (b) the application of magnetic field stabilized the suppressed current state. However, the suppressed current state in the latter cases was in highly stable pA range. The difference in OMCs response can be attributed to the difference in the number of molecules and the leakage channels via the AlOx of a MTJ. If the current suppression was due to any mechanical damage or artifact then it is important to answer the following two questions: (a) why did high current state occur intermittently (Fig. 5 and Fig. 6b)? (b) Why did magnetic field respond to a damaged junction, that to in a consistent manner? It is noteworthy that effect of magnetic field on transport was studied at room temperature.

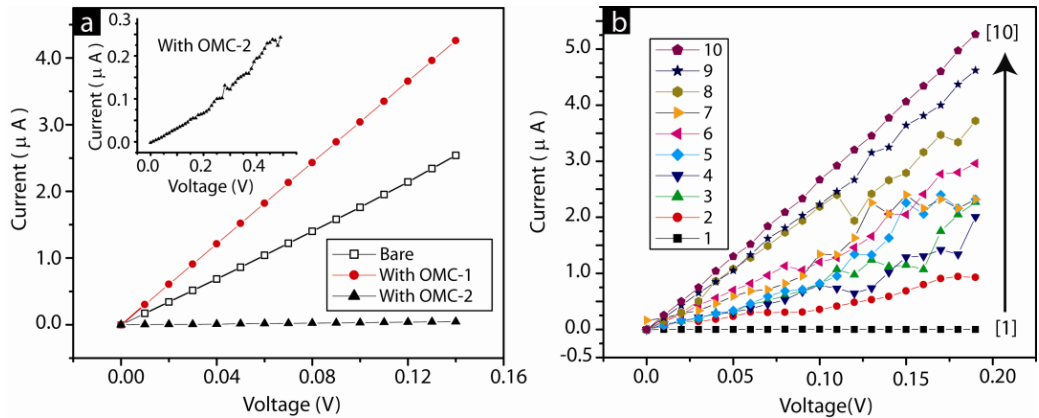

Fig. 7: Fourth MEMSD showing current suppression: (a) OMCs first increased the current and then a low current state stabilized; (b) repeating I-V gradually produced the high current state.

To further strengthen the observation of current suppression the effect of other factors were studied. For this purpose the effect of light radiation and the effect of repeating I-V studies, during the initial stage of current suppression, was investigated. On the fourth MEMSD, OMCs first increased the current and then within a few hours overall MEMSD's current decreased by more than 10 folds. Interestingly, a low current state commenced when sample was stored for few hours in a nitrogen ambience (Fig. 7a) at room temperature. On this sample, low current state (0.1 μA) gradually shifted to high current state when I-V measurements were repeated (Fig. 7b). This observation is consistent with the stabilization of a high current state with the repeating of I-V studies, after the observation of suppressed current state during the second MR study (Fig. 5b).

After the stabilization of a low current state on the fourth MEMSD, MR study was performed. Additionally, this MEMSD sample was periodically irradiated with white light from FL3000 Micro-Lite illuminator. Light radiation during MR study produced a transient high current state (Fig. 8a). Current cycled between nA to μA range. The observation of such a dramatic current change on the same MEMSD is not possible due to any mechanical damage; in the event of mechanical damage only a permanent suppressed current state was expected. After this MR study, the fourth MEMSD again stabilized into a suppressed current state. Interestingly, periodic light radiation in the absence of magnetic field appeared to stabilize a higher current state (Fig. 8b). Typically, the current state on freshly OMCs treated MTJ (Co/NiFe/AlOx/NiFe) or the MEMSD remained highly unstable for several hours. In the stable suppressed current state, this MEMSD produced a clear photoresponse. White light radiation produced ~400% change in MEMSD's current (Fig. 8c). To investigate the effect of magnetic field on the photoresponse, this fourth MEMSD was subjected to in-plane magnetic field. Application of the magnetic field changed the direction of photoresponse (Fig. 8d). Interestingly, the dark current kept moving towards a higher current state with time. These experiments (Fig. 8) emphasized that the light

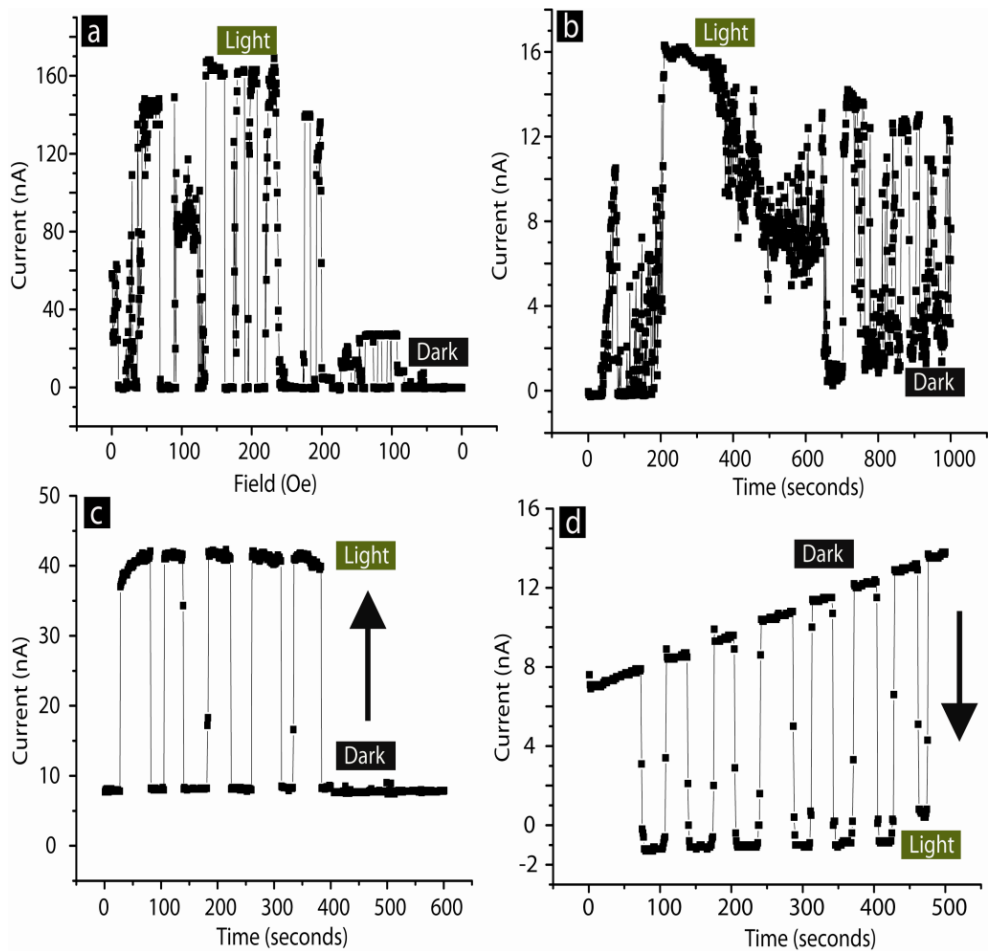

Fig. 8: Effect of light and magnetic field on the suppressed current state: (a) suppressed current state attempting to settle in high current state when sample was irradiated during MR study, (b) effect of light radiation brought the high current state with higher frequency. (c) MEMSD's photo response in the suppressed current state. (d) The direction of photoresponse changed after the in plane magnetization with 0.45 T. Current in panels (a-d) were measured at 50 mV

and magnetic field both influenced the transport through the MEMSDs. *Most of the MEMSDs, which showed current suppression, also exhibited the photovoltaic response.* A detailed discussion about MEMSD's photoresponse is presented elsewhere.

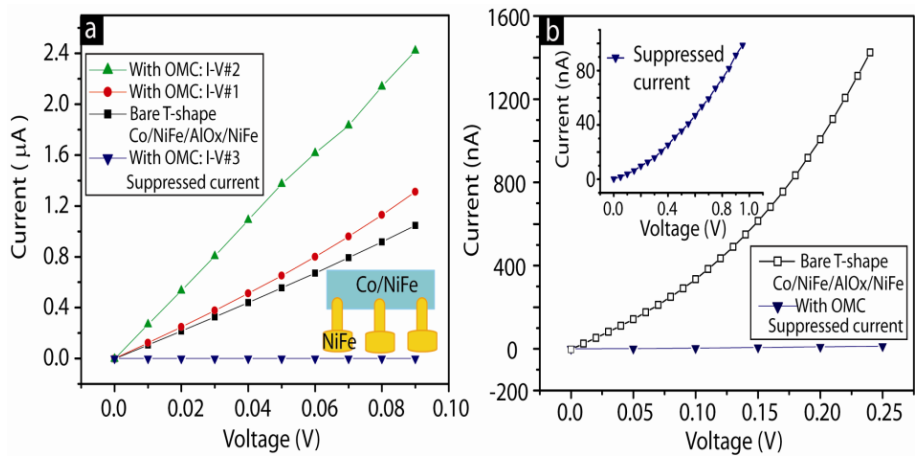

Fig. 9: Current suppression with isolated MEMSD: (a) on isolated Co/NiFe/AlOx/NiFe MTJ OMCs first increased the overall current level, and eventually a suppressed current state stabilized. (b) Suppressed current state was stable up to 1.0 V.

All the MEMSDs discussed so far were positioned on the separately processed chips. A typical chip contained 6x6 grid of cross bar form of MTJs. A crossbar junction array form has been considered highly promising for integrating spintronics devices [22, 26, 27]; such a device configuration can produce a highly defect-tolerant massive computational system, especially for the molecule based devices [28-31]. However, it is also likely that crosstalk among junctions can also lead to a strange and unpredictable I-V profile. To ensure that the current suppression on our MEMSDs did not arise from the interference due to the neighboring MTJs the isolated form of MEMSD was investigated. For this experiment isolated MTJs were prepared by depositing a 5 mm$^2$ Co/NiFe bottom electrode and then depositing a ~2 nm AlOx and 10 nm NiFe films, respectively, through the common photoresist cavity (inset of Fig. 9a). In essence, the exposed side edges for OMCs attachment were realized after the liftoff of photoresist. This protocol is exactly same as the one used for the crossbar device architecture for the discussed MEMSDs (Fig. 3-8). Ten isolated MTJs were fabricated on each chip (schematic of a section of chip with 3 isolated MEMSDs is shown in the inset of Fig. 9a). The isolated MEMSD design gave low yield; however, it is highly robust against the possible artifacts and damages. The major advantage of this approach is that a broad bottom electrode is unlikely to be discontinuous due to localized damages on the bottom electrode. The OMC bridges on the isolated Co/NiFe/AlOx/NiFe MTJ produced the similar current suppression phenomenon as observed with the same MTJ configuration in the crossbar forms (Fig.5-7). For the isolated MTJs discussed in figure 9, OMCs first produced the temporary current increase (Fig. 9a). After few hours device settled in the suppressed current state (Fig. 9a-b). The current suppression was highly stable and could not be influenced by the application of the bias up to ~ 1V (inset of Fig. 9b).

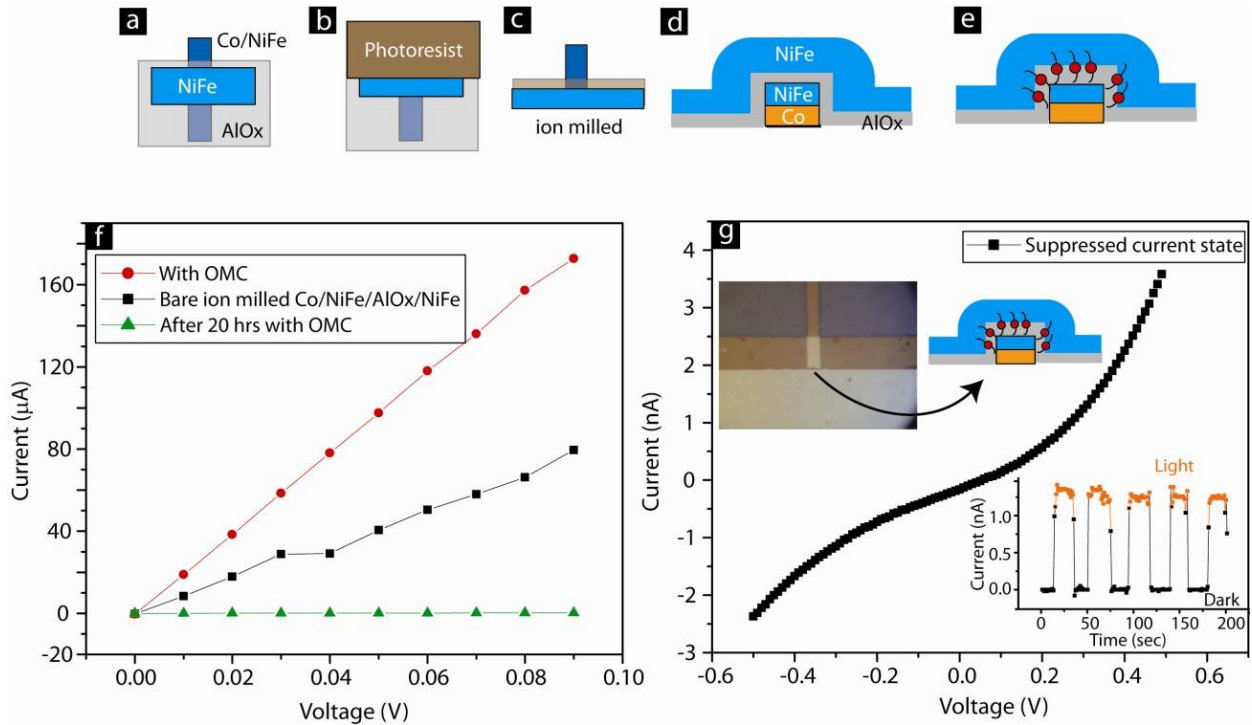

Fig. 10: Current suppression on ionmilled MEMSD: Fabrication of ion milled MTJ: (a) fabrication of a complete Co (5 nm)/NiFe (5 nm)/AlOx (2 nm)/NiFe (10 nm) MTJ, (b) creation creation of photoresist protection to selectively etch a part of MTJ, (c) ionmilling and liftoff of photoresist protection produced exposed edge MTJ. Side view of the ionmilled edge, (d) before and (e) after the bridging of OMC channels. (f) I-V data showing OMCs induced current increase and then current suppression, (g) OMCs induced suppressed current state also showed photovoltaic effect, inset showed the optical micrograph of 4 x 12 $\mu m^2$ MTJ and the schematic of its side view.

We also explored the likelihood of a systematic error, which could potentially result a suppressed current state on MTJs. In MEMSDs, a potential source of systematic error was expected from the *exposed edge creation step by the liftoff approach* (Fig. 1f-g). To investigate the possibility of liftoff-associated artifacts, liftoff based approach was replaced with an alternative method for creating exposed side edges. In the alternative approach, an ionmilling step was utilized to produce exposed side edge for the bridging of OMC channels. Fabrication of the ionmilling based MEMSD started with the fabrication of a Co/NiFe/AlOx/NiFe MTJ; here, AlOx was deposited in a larger area (Fig. 10 a). In the next step a part of the cross bar MTJ was covered with a photoresist protection (Fig. 10b). Next, ionmilling step was performed to etch away the unprotected section of the MTJ. After ionmilling, liftoff of the photoresist protection yielded a MTJ with one exposed side edge (Fig. 10c). Side view of the MTJ from the direction of ionmilling is shown in figure 10 d. Finally, OMC channels were bridged across the AlOx insulator to complete the MEMSD fabrication (Fig. 10e).

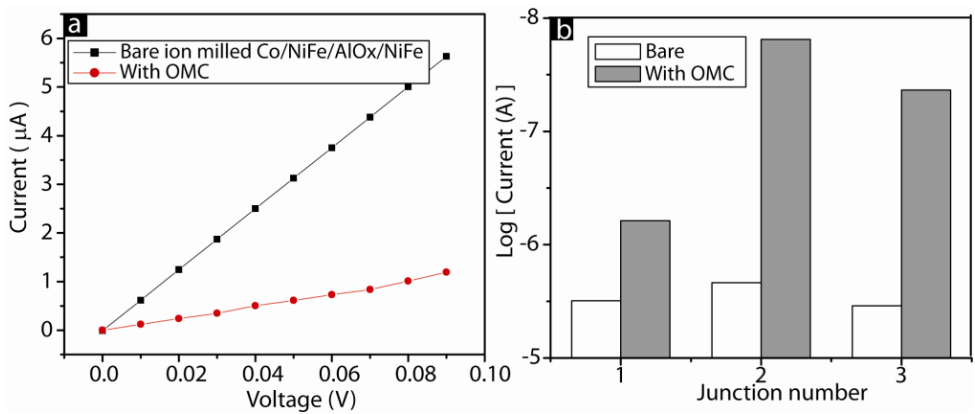

Fig. 11: Current suppression on unstable ionmilled MTJ: (a) OMCs lowered current of a MTJ with high leakage current. (b) OMCs induced transient current suppression on different MTJs of the same chip.

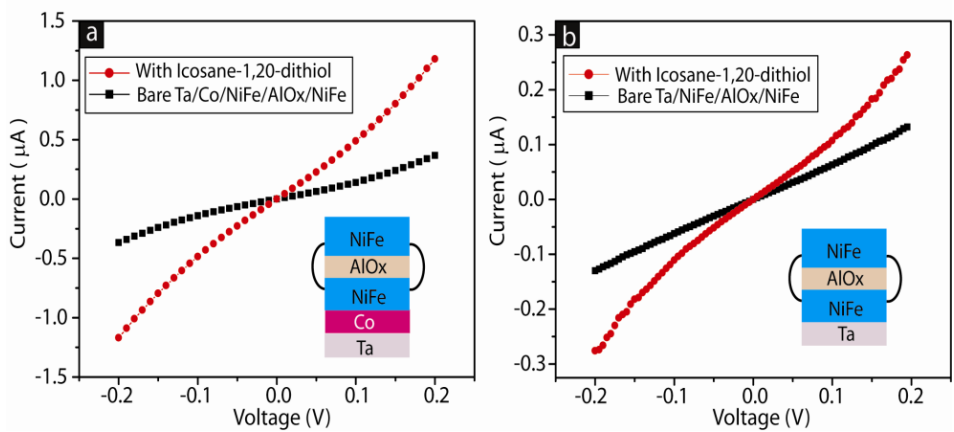

Fig. 12: Effect of alkane molecules: Bridging of alkane (icosane-1, 20 dithiol) molecular channels across AlOx of (a) Ta/Co/NiFe/AlOx/NiFe and (b) Ta/NiFe/AlOx/NiFe tunnel junctions.

The OMCs changed the transport characteristics of the ionmilled MTJ. OMCs first temporarily increased the current; after few hours, OMCs treated ionmilled MTJ settled in the suppressed current state (Fig. 10f). Here, the OMCs induced current increase and then the commencement of the suppressed current state is consistent with the results observed on liftoff produced MTJs (Fig. 6, Fig. 7, and Fig. 9). In the suppressed current state the ionmilled MEMSD also exhibited a photovoltaic effect (lower inset of Fig. 10 b), which is consistent with the same photovoltaic phenomenon observed with the liftoff based MEMSD (Fig. 8). The two important implications of this study were: (a) liftoff approach did not produce a systematic error and (b) multiple MTJs in the crossbar design did not create any crosstalk-induced artifacts leading to current suppression.

Ionmilled MTJs were highly useful to perform control studies. However, due to stress induced instability most of the ionmilled MTJs were short lived [20]. Unstable MTJs generally ended up in a higher current state (μA range) due to the failing tunnel barrier (Fig.11 a). Inspite of instability, a transient current suppression was also observed on several MTJs (Fig. 11b). Simultaneously processed three unstable ionmilled MTJs showed the different extent of current suppression (Fig. 11b).

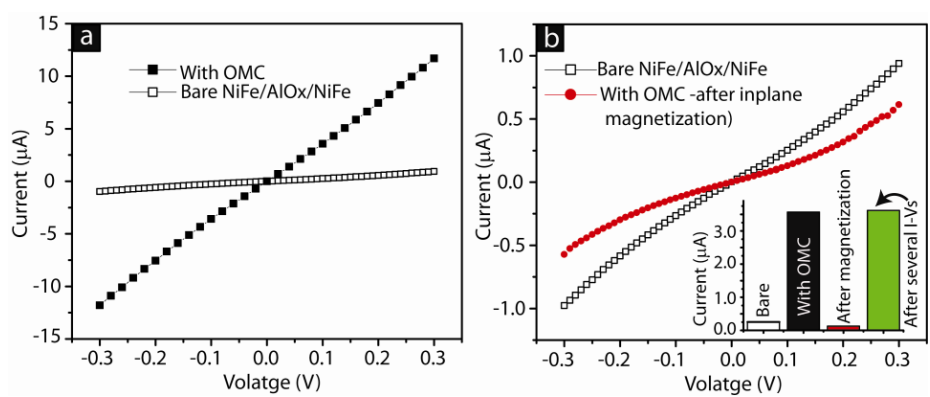

Fig. 13: Effect of OMCs molecules on NiFe/AlOx/NiFe: (a) OMCs generally increased the tunnel junction current, and (b) magnetization promoted moderately suppressed current state.

Effect of molecule type on the occurrence of current suppression was explored. *Are OMCs or molecules with a net spin state necessary for the observation of current suppression?* To investigate the role of molecules's spin, the simple alkane molecular channels with the thiol functional ends (Icosane-1,20, dithiol) were utilized. Icosane-1,20, dithiol molecules were synthesized by following the protocol established by Nakamura et al. [32]. The electrochemical molecule attachment protocol, as previously used for the OMC channels, was performed with a 5 mM icosane-1, 20, dithiol solution in di chloro methane solvent. Alkane molecules clearly enhanced the MTJ's current (Fig. 12a), and no current suppression was observed. These experiments also suggest an important point to support the observation of current suppression. If electrochemical step induced damages, not the OMC channels, yielded the current suppression then Icosane-1,20, dithiol molecular solution could also produce the current suppression, because exactly the same molecule attachment protocol was utilized. In addition to Co/NiFe/AlOx/NiFe another configuration of MTJ (Ta/NiFe/AlOx/NiFe) was also treated with Icosane-1,20 dithiol molecules. Icosane-1,20, dithiol molecular channels increased the MTJ current in the latter case too (Fig. 12b). Both results (Fig. 12a-b) confirmed that Icosane-1,20, dithiol molecules could establish the dominant conduction channels. This study suggested that a quantum dot like molecular channel with a net spin state is crucial for the observation of current suppression.

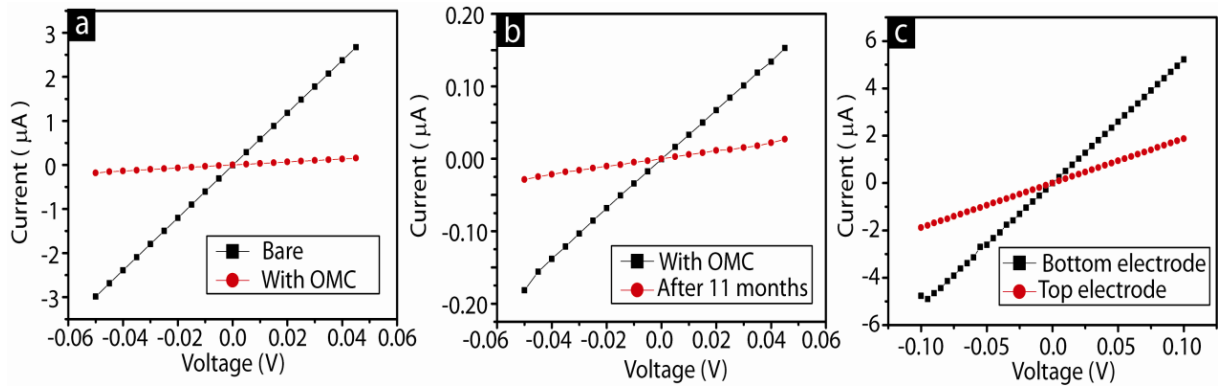

Fig. 14: Effect of OMCs on Ta/Co/NiFe/AlOx/NiFe/Ta: (a) OMCs reduced current on fresh MTJ, (b) this MEMSD after 11 months showed further current reduction. (c) Top and bottom electrode were in sound condition and current through them was in µA range.

The current suppression phenomenon was mainly observed with the Co/NiFe/AlOx/NiFe MTJ configuration. This configuration utilized two FM electrodes with different magnetic properties. According to our magnetization and ferromagnetic resonance (FMR) studies Co/NiFe and NiFe FM electrodes have dissimilar saturation fields and spin wave dynamics [33]. In the simple term this MTJ configuration possessed one hard ferromagnet (Co/NiFe) and one soft ferromagnet (NiFe). Such a combination of hard and soft ferromagnet was also found to be crucial for the observation of Kondo resonance splitting, observed with $C_{60}$ molecule in Ni break junction based MSD [14]. *Is this combination of hard and soft FM electrodes crucial for the observation of current suppression?* In an analogous system nature of magnetic interaction between iron nanoclusters, *analog of OMCs*, and FM electrodes of different magnetic hardness was investigated. Magnetization studies exhibited that nanoclusters interacted differently with the FM electrodes of different magnetic hardness [24], and also enhanced the inter-electrode exchange coupling in a MTJ.

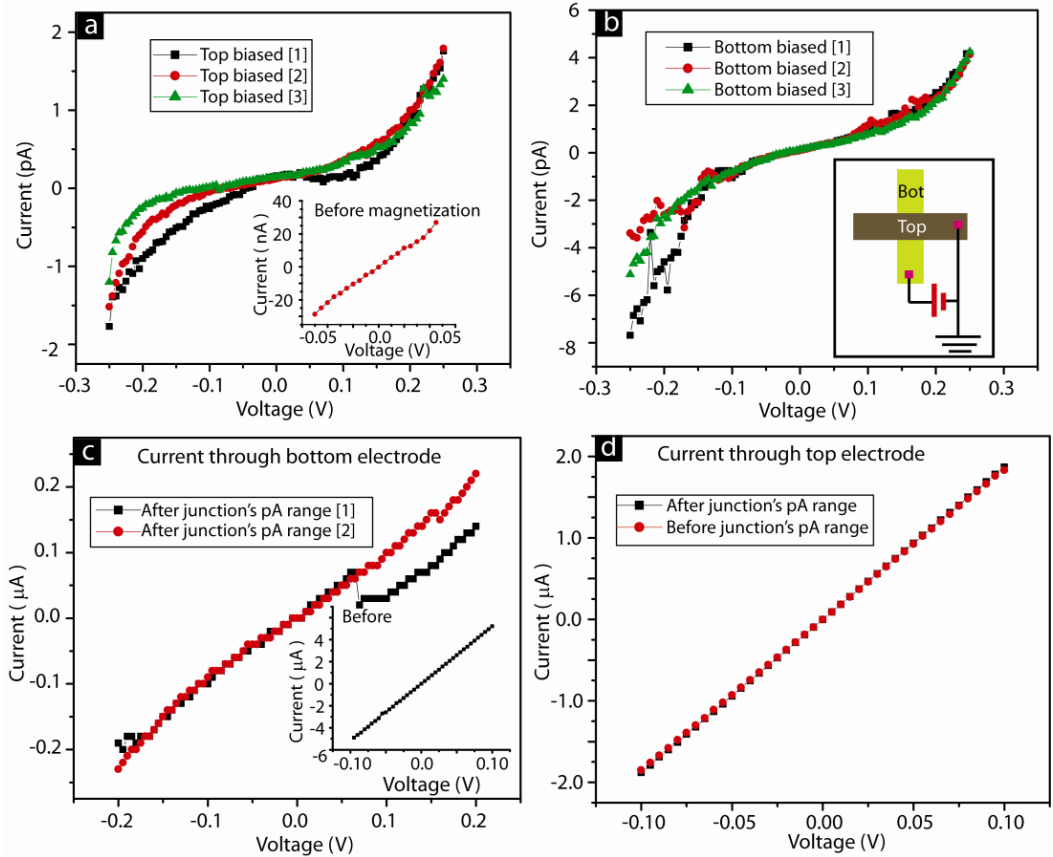

Fig. 15: Magnetization induced current suppression: (a) in-plane magnetization in 0.45 T field produced suppressed current state in pA range; (b) reversing the bias direction did not influence the suppressed current state. I-V study through (c) bottom electrode and (d) top electrode.

To ascertain that current suppression required interaction of OMCs with FM electrodes of different magnetic hardness complementary experiments were performed. For such studies the MEMSD were fabricated with NiFe/AlOx/NiFe MTJ test beds, employing identical FM electrodes of similar magnetic hardness. Generally, OMCs increased the MTJ's current (Fig. 13a). However, the NiFe (10 nm)/AlOx (2 nm)/NiFe (10 nm) MTJs where widths of the top and bottom electrodes were different from each other a transient low current state was produced by OMCs. It is presumably due to the difference in magnetic hardness of the two electrodes of differing dimensions. Previously, different dimensions of Ni electrode were utilized to produce the FM electrodes of different magnetic hardness for a breakjunction based MSD[14]. In the present case, a NiFe (10 nm)/AlOx (2 nm)/NiFe (10 nm) MTJ with 3 μm wide top electrode and 6 μm wide bottom electrode exhibited an OMC induced high current state (Fig. 13a). When this MEMSD was magnetized in in-plane magnetic field the MEMSD current became even lower than the bare tunnel junction current (Fig. 13b). The magnetization induced low current state is consistent with the observations of similar phenomenon on MEMSD with Co/NiFe/AlOx/NiFe configuration (Fig. 5 and Fig. 6). However, OMCs on Co/NiFe/AlOx/NiFe MTJ produced a much stronger current suppression. Additional control experiments were performed to ensure that electrochemical step for OMCs attachment do not etch FM electrodes or create any damage leading to current suppression. In the first attempt a large number of MTJs were prepared with the ~4.0 nm thick AlOx barrier, so that ~3.0 nm long OMCs were unable to bridge this AlOx gap to cause any current suppression. In this situation any defect/artifact producing current suppression could be easily detected. Regularly used electrochemical OMC attachment protocol was applied on these control samples to verify if this step caused particular type of damage leading to current suppression. None of these sample exhibited current suppression or current increase.

To further investigate that any serendipitous damage to electrical leads is not the reason behind current suppression, the following control experiment was performed. For this experiment Ta/Co/NiFe/AlOx/NiFe/Ta MTJs were utilized. In bottom electrode a Ta layer under Co served as an adhesion layer and minimized the possibility of localized damage to the bottom electrode due to deadhesion or accidental etching of Co/NiFe. In the top electrode the Ta above the FM electrode ensured that NiFe is not oxidized during the course of study; it is noteworthy that NiFe by itself was highly stable against electrochemical step and OMC solution. Addition of Ta in the top and bottom electrode enhanced the MTJ's durability. A number of samples studied with this configuration also showed several order change in MEMSD transport, as observed with Ta/Co/NiFe/AlOx/NiFe MTJs. The aims of present control experiment were: (a) check the integrity of electrical leads in the suppressed current state (b) attempt to reverse the current suppression by forcing current in the opposite direction. The second aim was mainly to investigate the mechanism behind the current suppression.

This crucial control experiment was performed with Ta/Co/NiFe/AlOx/NiFe/Ta MTJ configuration. OMCs bridges across insulator produced a suppressed current state (Fig. 14a); MTJ's current reduced by one order. After initial I-V studies this MEMSD sample was stored in the nitrogen ambience at the room temperature. Restudying this sample after 11 months showed that the junction was still in the suppressed current state; moreover, 11 months incubation further reduced the MEMSD current (Fig. 14 b). It is noteworthy that the top and bottom electrode were intact and current through them was in µA range (Fig. 14 c), the typical current magnitude observed with freshly deposited top and bottom magnetic electrodes was also in µA range. After checking the integrity and electrical continuity of the top and bottom electrodes, attempts were made to achieve the higher degree of the suppressed current state. According to our experience with Co/NiFe/AlOx/NiFe based MEMSDs in-plane magnetization promoted the highly suppressed current state. In the present control experiment on MEMSD with Ta top layer, the application of 0.45 T static in-plane magnetic fields produced a suppressed current state at room temperature (Fig. 15a). This in-plane magnetization caused a stable current reduction by more than four orders of magnitude. We also utilized this experiment to probe the mechanism of current suppression in MEMSDs. For instance, we explored the effect of direction of current flow on the MEMSD's current state. To investigate if reversing the current flow can disrupt the suppressed current state bottom electrode was biased (inset of Fig. 15b); for the I-V studies in figure 15a the top electrode was biased. Reversing the current flow direction produced insignificant changes in the suppressed current state. It is noteworthy that during first few hours after OMCs attachment, when a MEMSD keep changing current magnitude dramatically, sometime changing the bias direction affected the MEMSD transport, but in an irreproducible manner. Next, I-V studies were performed through the top and bottom electrodes when the junction was still in the suppressed current state. It is noteworthy that during I-V studies the electrical probes touched NiFe (Ta) layer of Ta/Co/NiFe (NiFe/Ta) bottom (top) electrode. Interestingly, the magnitude of current through the bottom electrode reduced by a factor of 40 (Fig. 15c). However, the magnitude of bottom electrode current was still ~seven orders higher than that of MEMSD junction. *Can this change in bottom electrode current be due to the evolution of a new magnetic ordering in the molecular junction area?* According to previous studies spin transport in FM films is significantly influenced by the spin scattering at the boundary of magnetic domains with differing spins [27, 34]. As discussed elsewhere in this study, OMCs induced dramatic changes in the magnetization and spin wave dynamics of the MTJs used in this study [33]. The transport through the top electrode was also studied. I-V response of the top NiFe/Ta electrode remained unchanged (Fig. 15d). During the study of top electrode, electrical probes were in the contact of nonmagnetic Ta layer, which is presumably indifferent from the probable upheaval in the magnetic ordering in MTJ's vicinity. However, it is not clear if NiFe of the top electrode also underwent any magnetic reordering.

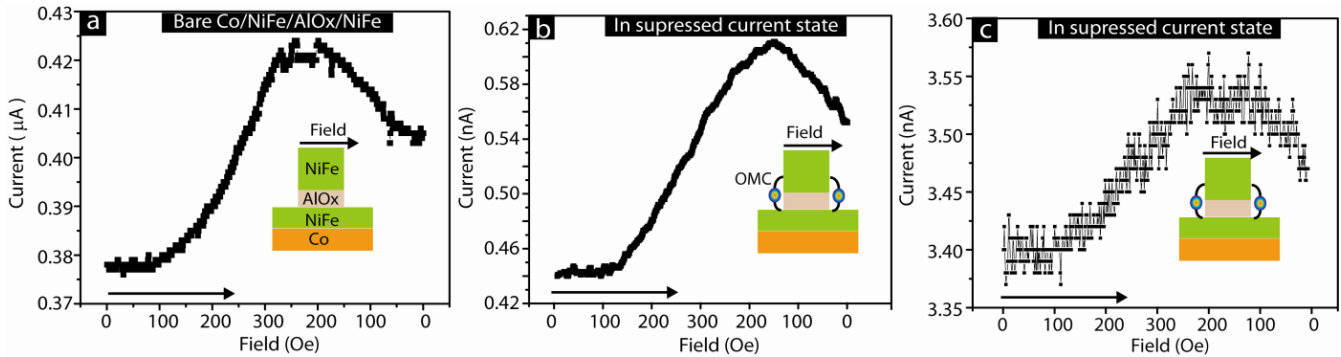

Fig. 16: Effect of magnetic field on MEMSD current: (a) current variation of bare MTJ with magnetic field. (b) and (c) panels shows the current variation of MEMSDs in the suppressed current state.

To understand the mechanism behind current suppression effect of magnetic field was investigated on spin transport. The magnetic field influenced the spin transport of Co/NiFe/AlOx/NiFe based MEMSD. In the first study the current through the MEMSD device was continuously measured with the variation in magnetic field. The inplane magnetic field was first increased to 300 Oe and then decreased back to 0 Oe. The variation in the MTJ resistance with magnetic field is shown in figure 16a. For this study, bare MTJ current was studied at 40 mV. MTJ's current kept increasing gradually even past the maximum applied field (300 Oe) and then eventually decreased to a level higher than that observed in the beginning of this study. After the OMCs attachment, the current through the MEMSD stabilized in the suppressed state. MR studies in the suppressed current state closely resembled with the typical MR response on bare MTJ samples (Fig. 16b-c). MEMSD with Ta/Co/NiFe/AlOx/NiFe MTJ did not show abrupt and instantaneous switching with magnetic field.

Influence of the stronger magnetic field was also studied on MEMSD's transport at room temperature. MEMSD's magnetization in a 0.45 T static inplane magnetic field for 0.5-2 hours further suppressed the current. Magnetization decreased the MEMSD's current by more than three orders (Fig. 17). However, static magnetic field induced suppressed current state slowly moved to a higher current state (Fig. 17a). The recovery phase was generally expedited by the multiple I-V studies (Fig. 7). Interestingly, recovered current state has similar profile as observed prior to the magnetization of this MEMSD (Fig. 17a); however, the magnitude of MEMSD current was significantly different before and after magnetization. The magnetization induced current suppression is in agreement with the similar phenomenon observed with other MEMSDs (Fig. 4-6).

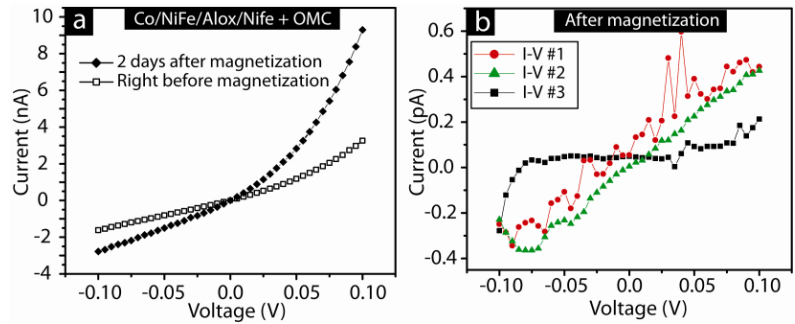

Fig. 17: Effect of magnetization on MEMSD: (a) MEMSDs before and after the commencement of magnetization induced suppressed current state (b) I-V in suppressed current state.

The effect of magnetic field was also studied on the slightly different MTJ configuration. For one MEMSD configuration, the top NiFe electrode was covered with the Ta metal. Surprisingly, the deposition of Ta changed the nature of inter-electrode coupling from weakly antiferromagnetic to weakly ferromagnetic; our FMR studies, discussed elsewhere [33] and briefly in this paper, confirmed the Ta effect on the nature of magnetic coupling. A MEMSD with Ta on top exhibited a current lowering due to OMCs (Fig. 18a). One of such samples repeatedly showed switching between μA to pA current state within the first few hours after OMCs attachment. During MR studies, a MEMSD with the Ta on top started switching its current state from suppressed to μA current range. During MR studies, the top electrode was biased at 50 mV and bottom electrode was grounded; the strength of in plane magnetic field, along with the top electrode direction, was increased from 0 to 200 Oe and then reduced to 0 Oe. In the present case, multiple switching between current states occurred during the MR study at room temperature (Fig. 18 b). Initially, MR1 and MR2, did not show the switching and current persist in pA range. From MR3 onwards, switching occurred in three consecutive MR studies. In MR3 current increased from pA to μA for a brief duration (Fig. 18b). In MR4 current started in μA range and as magnetic field increased, MEMSD transcended to a suppressed current state. During MR5 current again started from μA level and after staying there for a longer period it finally moved to the suppressed current state with increasing magnetic field. Various current plateaus observed during MR4 and MR5 are plotted (Fig. 18c). During the MR6, current remained in μA range at all the time. It is noteworthy that during MR3-5 the application of 50 mV appears to promote the high current state. Previously, repeating I-V studies switched the MEMSD's suppressed current state to higher one (Fig. 4 and Fig.7). It is also noteworthy that the application of magnetic field promoted the suppressed current state; this observation is consistent with the magnetic field induced suppressed current state on most of the MEMSDs (Fig. 4-6, Fig.3, and Fig. 15).

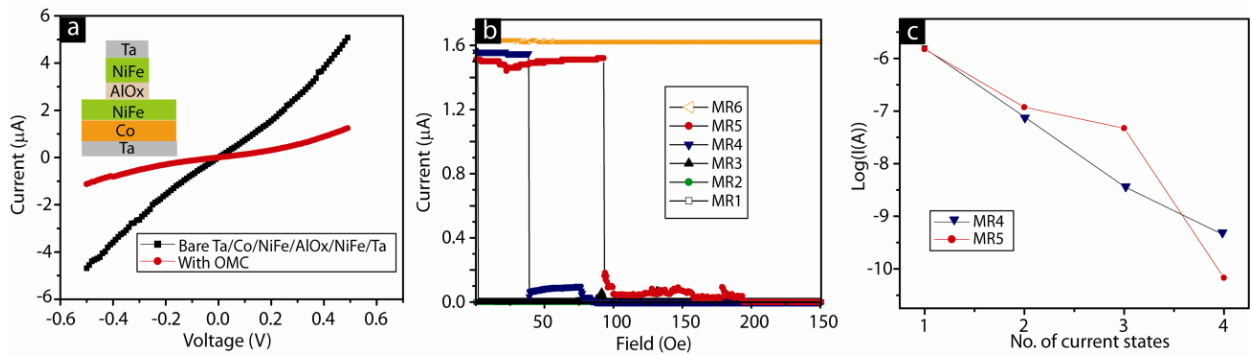

Fig. 18 Seven orders current switching on Ta/Co/NiFe/AlOx/NiFe/Ta based MEMSD: (a) OMC stabilized a lower current state. (b) Magneto resistance (MR) studies showing current state switching between μA and the suppressed current state. (c) Observed current plateaus during MR 4 and 5; line connecting data point is only guide to eye.

One important aspect of switching from μA to suppressed current state during MR3-5 is that a number of intermediate current states were also observed (Fig. 18c). The observation of intermediate current states signifies that switching between current states was not due to serendipitous loss of electrical connections. It is also noteworthy that the μA to pA level change on a fresh MEMSD with Ta on top FM electrode (Fig. 18) is consistent with the 11 month old MEMSD with the same configuration (Fig. 14-15). We also observed that a MEMSD with Ta/NiFe top electrode was more easily switched between the μA and suppressed current state than that observed with the Co/NiFe/AlOx/NiFe. According to our magnetic measurements, as discussed elsewhere in this study, the magnetic properties of a MEMSD with NiFe/Ta were significantly different from that of MEMSD with the NiFe top electrode. Several MEMSD with NiFe top also showed dramatic abrupt switching between current states (Fig. 5, 6, and 8). However, we could not control the switching at will for the studied MTJ and OMCs systems. We envision that a suitable combination of molecules and MTJ can allow controllable switching between different current states differing by several orders of magnitude, and hence leading to ultimate spin valves.

It is noteworthy that the ~7 orders current change in a molecular spintronics device has already been predicted. According to Petrov et al. [10] "In the case of an antiferromagnetically ordered quantum molecular wire, the inter-electrode tunnel current can be regulated by a magnetic field in a wide region up to 7-8 orders of magnitude because of the ability of the magnetic field to influence changes in the spin orientations". We experimentally observed a 3-6 orders change in the current magnitude of MEMSDs' current. However, mechanism behind switching on MEMSD is quite different and requires additional insightful experiments.

The effect of magnetization in the static magnetic field was also studied on the MEMSD with the *Ta/NiFe top electrode*. A 0.45 T in plane magnetic field suppressed the MEMSD current by more than five orders during the period of magnetization (Fig. 19 a). This observation is in agreement with the magnetization induced current suppression observed on the MEMSD with only NiFe top electrode (Fig. 17), after the magnetization under identical circumstances. However, difference between the current states before and after magnetization was nearly two orders higher for the latter case (Fig. 18). Interestingly, current through the bottom magnetic electrode (Ta/Co/NiFe) also changed by more than 10 folds after the magnetization step (Fig. 19 b). *The change in electrode resistance along the junction resistance is indicative of the significant change in the magnetic ordering*. Unfortunately, due to the lack of unhampered experimental resources we could not perform extensive transport studies on MEMSD's FM electrodes in the suppressed current state (Fig. 15). Transport studies of the FM electrodes are strongly recommended in future MEMSDs, for the validation of the present work and to further understand the MEMSD system.

Dramatic changes in the MEMSD transport with the magnetic field prompted us to investigate the magnetic properties of a MTJ+OMCs system or MEMSD. Initially, the MEMSD showing current suppression was subjected to magnetic force microscopy (MFM) study. MFM studies were performed using a Digital Instrument Multimode AFM and Molecular Imaging Pico Scan AFM. Highly sensitive MFM tips (supplied by Nanosensors) with the following specifications were utilized: type PPP-MFMR, tip side Co coated, force constant in 0.5-9.5 N/m range, and resonance frequency 45-115 kHz. During MFM scans the gap between the tip and the substrate was 10-150 nm. For the high resolution and uninterrupted MFM studies a 100 nm tip sample gap was maintained. The MFM study of the bare MTJ exhibited an insignificant difference in the top and bottom electrodes (Fig. 20a). However, the MFM studies of MEMSD's top and bottom electrodes exhibited an unusually large contrast (Fig. 20b). Now, there can be an apprehension that MFM contrast on the MEMSD is caused by the serendipitous experimental settings (Fig. 20b). However, such a contrast difference was also observed on the two more MEMSDs. Further support for these MFM studies is presented in the next chapter. The MFM studies in figure 20 exhibited that OMCs effect was not limited in the junction area only. *Due to this reason, we were unable to estimate the true energy conversion efficiency of the photovoltaic energy produced on MEMSDs (Fig. 8 and 10)*. This MFM study also indicated that the dramatic current suppression is correlated with the magnetic ordering of the FM electrodes. It is important to recall that the change in the magnetic contrast of the bottom FM electrode, as compared to the magnetic color of the top electrode, complement the observation of ~40 fold change in resistance when a MEMSD switched to the suppressed current state (Fig. 15).

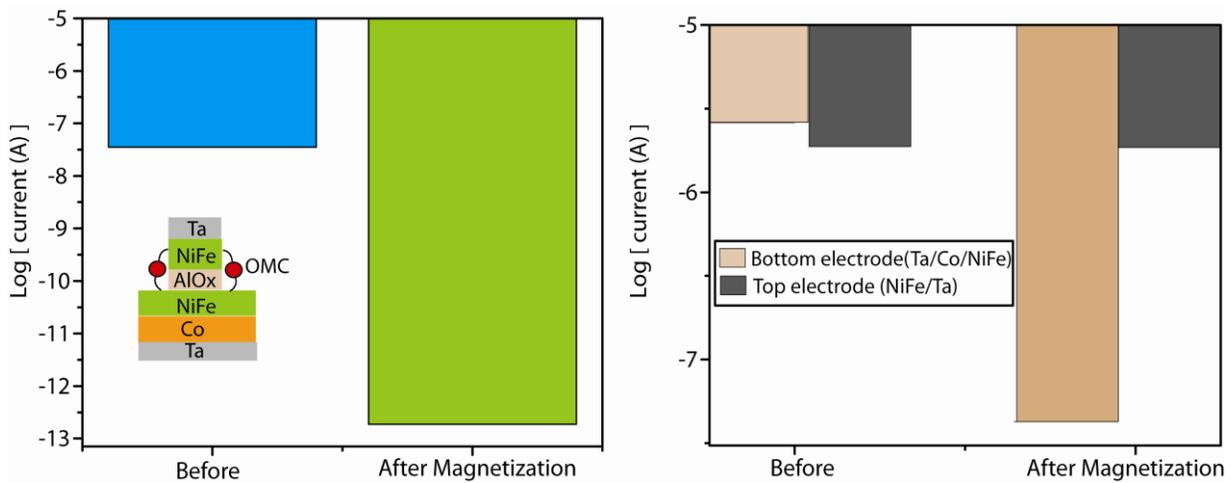

Fig. 19: Magnetization induced current suppression on MEMSD with top Ta layer: (a) magnetization inducing suppressed current state. (b) Effect of magnetization on transport through the top and bottom electrode.

To investigate the effect on OMCs induced magnetic changes, two other magnetic characterization strategies were employed. However, to magnify the effect of OMCs the MEMSD in the disk form, not the cross junction with extraneous FM electrode areas, were utilized. Schematic and optical micrograph of the actual dots is shown in the inset of Fig. 21a. Every sample contained 7000 to 21,000 MTJ dots. Firstly, the magnetization studies were conducted with Quantum design MPMS-XL-7T Squid magnetometer at 150 K. The OMCs transformed the magnetization of Ta/Co/NiFe/AlOx/NiFe; typical hysteresis curve became linear (Fig. 21a). A linear magnetization response indicates the development of antiferromagnetic coupling [35, 36]. Antiferromagnetic coupling on iron/silicon/iron MTJ with similar junction area produced the linear magnetization curve for the limited range of magnetic field [36]. The magnetic field where a linear magnetization curve saturates can be used for calculating the strength of magnetic coupling between the two FM electrodes according to the following expression.

*J (exchange coupling) = Saturation field x saturation magnetization x film thickness*

However, we could not observe the saturation of OMCs induced linear magnetization graph up to 3T. Could the lack of saturation be due to the emergence of a strong antiferromagnetic coupling? Before seeking answer to this question it is crucial to ensure that there is no other plausible reason for the same. At this point one may also question if the appearance of a linear magnetization after OMCs treatment is merely due to the artifact on accidently damaged sample. For instance, linear magnetization may appear due to the presence of paramagnetic materials [37] and defects within the MTJ [25]. However, any mechanical damage was not obvious when MEMSDs were studied under the optical microscope and AFM. Moreover, strong influence of OMCs was also observed on another MTJ configuration. OMCs increased magnetization of the MTJ with Ta/Co/NiFe/AlOx/NiFe/Ta configuration (Fig. 21b). Samples utilized in the two studies possessed the same MTJ dot density and dimensions; only difference was in the top electrode composition. This trivial appearing change in top electrode configuration of test bed MTJ was latter found to produce a pronounce effect on the nature of inter-layer exchange coupling between the two FM electrodes (Fig. 21c-d) [33].

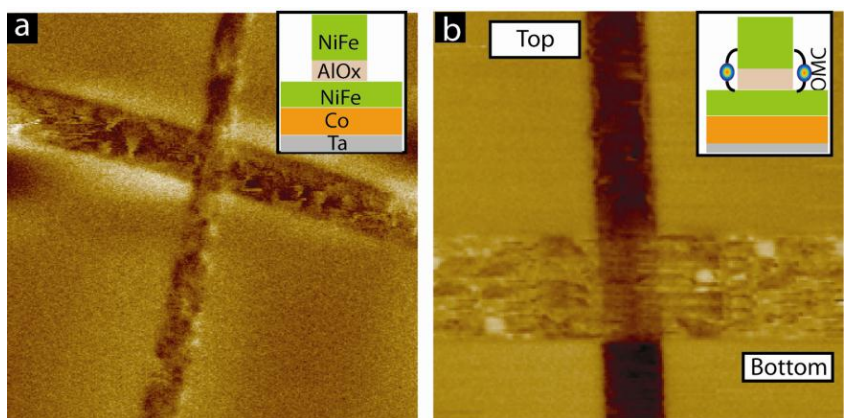

Fig. 20: MFM imaging of a Co/NiFe/AlOx/NiFe MTJ (a) before and (b) OMCs attachment.

To investigate the effect of Ta and OMCs on the magnetic properties of the bare MTJ ferromagnetic resonance (FMR) studies were conducted. For FMR studies Brucker EMX EPR spectrometer equipped with Brucker Mirowave Bridge ER 041MR and Brucker Power Supply ER 081(90/30) was utilized. For all the experiments, a ~9.7 GHz microwave frequency and the room temperature were used. A magnetic field was applied in the sample plane to study the uniform modes of the thin films and multilayers.

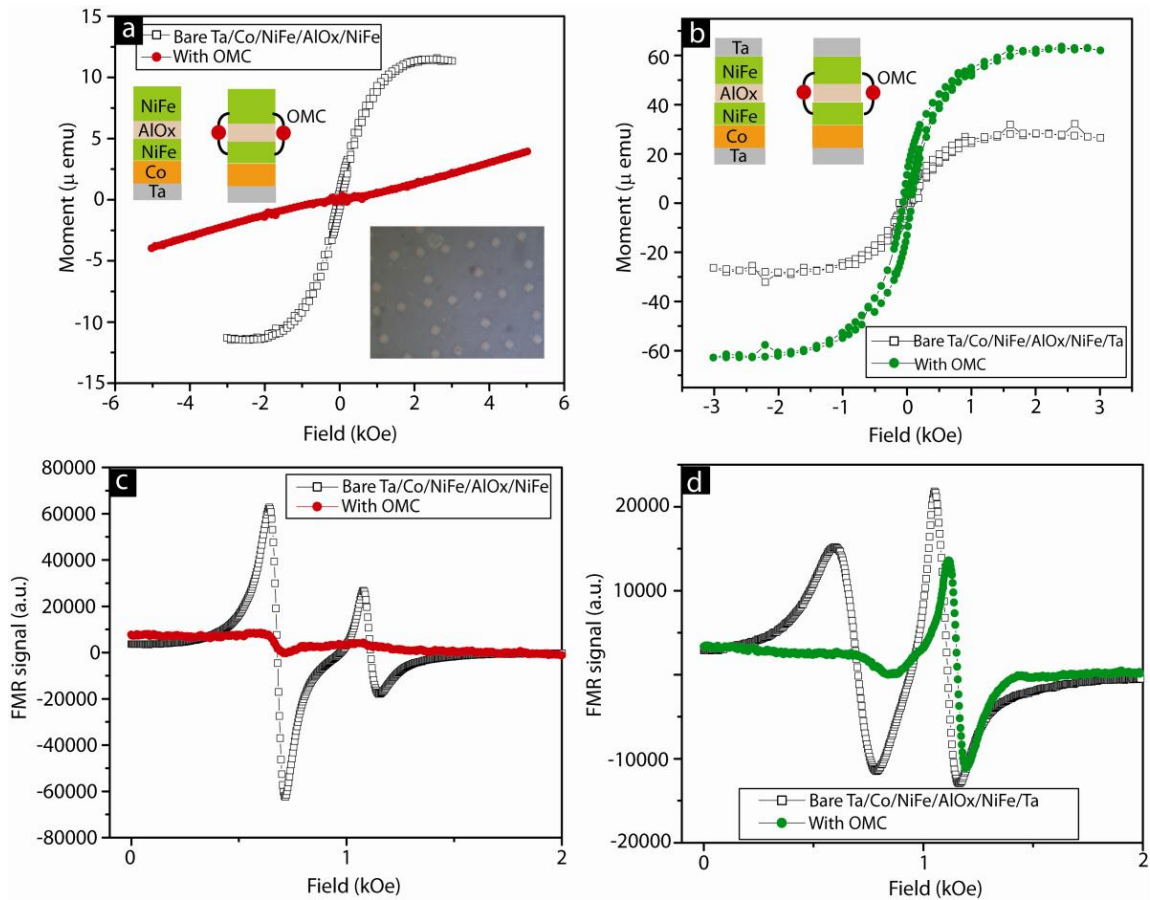

Fig. 21: Effect of OMCs on the MTJ's magnetic properties: (a) OMCs made magnetization of Ta/Co/NiFe/AlOx dots linear. (b) OMCs increased the magnetization of Ta/Co/NiFe/AlOx/NiFe/Ta. OMCs strongly influenced the FMR of (c) Ta/Co/NiFe/AlOx/NiFe and (d) Ta/Co/NiFe/AlOx/NiFe/Ta.

Bare MTJs with and without Ta top layer showed remarkable changes in the relative intensity of the two FMR modes. The relative intensities of the first and the second FMR modes for MTJ *without the Ta top layer* (Fig. 21c) and *with the Ta top layer* (Fig. 21d) corresponded to the weak antiferromagnetic and the ferromagnetic couplings, respectively [38]. OMCs on these two types of MTJ dots produced opposite effect, presumably because of the inherent difference in the nature of coupling, across the AlOx insulator. FMR produced other remarkable insights into OMCs effect. OMCs suppressed the FMR modes of Ta/Co/NiFe/AlOx/NiFe. According to FMR modeling for a ferromagnet-spacer-ferromagnet system such a mode disappearance occurred when the two FM electrodes are coupled via strong antiferromagnetic exchange coupling [39]. On the other hand, the FMR spectra of Ta/Co/NiFe/AlOx/NiFe/Ta configuration with OMC bridges exhibited single FMR mode (Fig. 21c). However, FMR spectra of a typical Ta/Co/NiFe/AlOx/NiFe/Ta MTJ showed two modes (Fig. 21d). According to FMR modeling the development of a single mode occurred when ferromagnetic coupling between the two FM electrodes became very strong [39]. The FMR and the magnetization study provided complimentary evidences in the support of our hypothesis that OMC induced dramatic changes in the magnetic properties of FM electrodes. These drastic changes in magnetic properties are expected to produce dramatic changes in the MEMSD transport, and presumably produced the current suppression like phenomenon. An extensive discussion on the OMC induced magnetic properties is furnished in the dedicated paper [33]. After the discussion on the experimental studies apropos of current suppression following is the likely mechanism behind the current suppression.

Following sections discuss the likely mechanisms behind current suppression observed with MEMSDs. MEMSDs are produced by attaching the OMCs across the insulator of a MTJ. A typical MTJ shows two resistance states; $R_P$ and $R_A$ resistance states occurs when the magnetizations of the two FM electrodes of MTJ are parallel and antiparallel to each other [21, 40]. Tunneling magneto resistance (TMR) is the key device characteristic of a MTJ and can be defined in terms of degree of spin polarization (P) of the individual FM electrodes:

$$TMR = \frac{R_A - R_P}{R_P} = \frac{P_1 P_2}{1 - P_1 P_2} \qquad (1)$$

Here $P_1$ and $P_2$ are the spin polarization of two FM electrodes. The spin polarization of a FM electrode is defined by the density of states (DOS) of up ($D_\uparrow$) and down ($D_\downarrow$) spins near the Fermi energy level ($E_F$). P of a FM electrode is dependent on the itinerant sp orbital electrons. These sp electrons travel longer distance, as compared to the localized d-orbital electrons which dominantly govern the magnetic moment and magnetization of FM electrodes [40].

$$P = \frac{D_\uparrow - D_\downarrow}{D_\uparrow + D_\downarrow} \qquad (2)$$

Here, we need $R_P$, $R_A$, and P for the MEMSD. These parameters are expected to be significantly different from the respective quantities for a MTJ test bed. According to our magnetic studies, OMCs stabilized antiferromagnetic coupling between the two FM electrodes of a MEMSD (Fig. 20). In this case, a MEMSD exhibited a suppressed current state, and presumably represented the $R_A$, with a magnitude of ~100 MΩ. The highest magnitude of current recorded on the MEMSD was in μA range around 100 mV, and this current state corresponded to $R_P$. $R_P$ in our studies was of the order of 0.1 MΩ. Using $R_P$ and $R_A$ of MEMSD the TMR magnitude was ~$10^3$ folds or ~$10^5$ %, according to eq.(1). Such a high TMR ratio is possible when P for both FM electrodes tend to be one [40]. It means, for FM electrodes of the MEMSD need to have single spin band around the $E_F$ (FM electrode's Fermi energy level). Now there are two key questions to focus on: (a) by what mechanism a regular NiFe FM electrode, with a typical P<0.5 [40], can became almost 100% spin polarized (P= ~1)? (b) Could the use of 100% spin polarized magnetic electrodes lead to such a large TMR? It is noteworthy that a number of materials possess single spin band near $E_F$ in the isolated state [40]. However, the utilization of even 100% spin polarized materials has not produced dramatic improvement in TMR ratio at room temperature [40]. Failure in seeing the large TMR ratio may be due to the reason that when such a 100% spin polarized materials are incorporated in a MTJ their $E_F$ started seeing bands of both type of spins. Other reason may be due to strong spin scattering within the insulator and at the interfaces of a MTJ [22]. Interestingly, our unusually high TMR ratio is observed with the FM films; we used Co/NiFe and NiFe FM electrodes.

For the FM electrodes P is generally found to be less than 0.5 [14, 27, 40]. *Most important fact about P is that it is not a material property* [40]. P strongly depends on the type and thickness of the spacer between the two FM electrodes [40]. P also strongly depends on the surface chemistry of the FM electrode [41]. For instance, the magnitude of P for Ni has been observed to vary from ~0.1 to ~0.4. It is crucial to focus on the method of calculating P and noticing following points: (a) P of a FM material is calculated by using a configuration of FM-insulator-superconductor devices; here superconductor served the role of a spin detector [40]. Interestingly, dramatic effect of enhancing exchange coupling was mainly observed on a system of FM-spacer-FM [14, 23, 42], which shared common features with MEMSDs. Now, it is important to investigate if P for the MEMSD can be entirely different. (b) P was measured by keeping the two electrodes, FM electrode and superconductor, separated by 1-2 nm thick metal oxide insulators. For 1-2 nm insulator thickness the wave functions of the two metal electrodes are weakly coupled to each other [43]. However, when molecule with a net spin state are present between the two FM electrodes a strong coupling is highly likely [14]; molecular devices frequently exhibited the signature of strong inter-electrode coupling [44]. Hence, in the context of MEMSDs or MSDs P must be calculated in the strong coupling regime. The calculation of P has never been performed for the strongly coupled FM electrodes; due to the lack of such important information, Pasupathy et al. [14] utilized the P calculated from a system which does not conform with their system showing anomalous Kondo splitting.

Can molecule induced exchange coupling produce dramatic change in magnetic properties of a FM electrode? We believe that MEMSDs represent the case when the two FM electrodes are strongly coupled by ~10,000 OMCs. Kondo resonance type signature of molecule induced strong inter-electrode coupling has been observed on break-junction based MSDs [44]. Kondo resonance results from the evolution of quasi states at zero bias when the localized spin on a molecular channel interacts with the itinerant spins on the metallic electrodes [44]. These quasi states lose their degeneracy when an external magnetic field is applied [45]. In a key MSD study, $C_{60}$ molecule(s) between the two Ni FM electrodes produced Kondo resonance splitting without the application of external magnetic field. More surprising was the extent of splitting (~15 mV), which was estimated to occur at equivalent external magnetic field of >50 T [14]. Need of such a strong magnetic field was fulfilled by the molecule induced strong exchange coupling producing a local magnetic field to produce observed Kondo splitting of ~15 mV [12-14]. *Now the question is whether a strong exchange coupling can affect the P of FM electrodes too.* P of a FM electrode result from the difference in spin DOS for the up and down electrons at $E_F$. To investigate the effect of OMCs induced coupling on P it is prudent to check if strong exchange coupling is capable of affecting the basic magnetic properties like Curie temperature, which depends on the spin DOS of FM electrodes. Experimentally, strong exchange coupling was found to change the Curie temperature of the FM electrodes by ~40 K when the thickness of the non-magnetic spacer between them was reduced to few monolayer [38]. Like the case of Kondo level splitting on a MSD [14], effect of exchange coupling on Curie temperature [23] was attributed to the spin fluctuations.

This discussion and future explaining of new MSD related phenomenon will immensely benefit from the direct theoretical and experimental attempts to measure the impact of molecule-induced exchange coupling on P and the magnetic properties of FM electrodes. Recent theoretical and experimental studies attempted to study the nature of magnetic interaction between the porphyrin magnetic molecules and the single Co FM electrode [46]. However, magnetic studies focusing on the simultaneous magnetic interactions of a magnetic molecule with the two FM electrodes can provide insight about the current suppression observed with MEMSDs. We have conducted magnetization and FMR study on MEMSD to garner further insights about the connection between the magnetic properties and the current suppression.

Our magnetic characterizations confirmed that OMCs affected the static and dynamic magnetic properties of a FM electrode (Fig. 21) employed in a MEMSD (Fig. 22). From these studies it is clear that ~10,0000 molecules on the MTJ edges can dominate the magnetic properties of a ~25 $\mu m^2$ area MTJ. In the separate studies, nanoscopic metallic shorts, analog to OMC channels on the MTJ's edges, were observed to transform the MTJ transport. Opets et al. [47] observed that only a few nm wide metallic short was sufficient to fail a tunnel junction of ~100 $\mu m^2$ area MTJ. By corollary, it is logical to think that ~10,000 OMCs can cumulatively dominate the planar area of a ~25 $\mu m^2$ area MTJ.

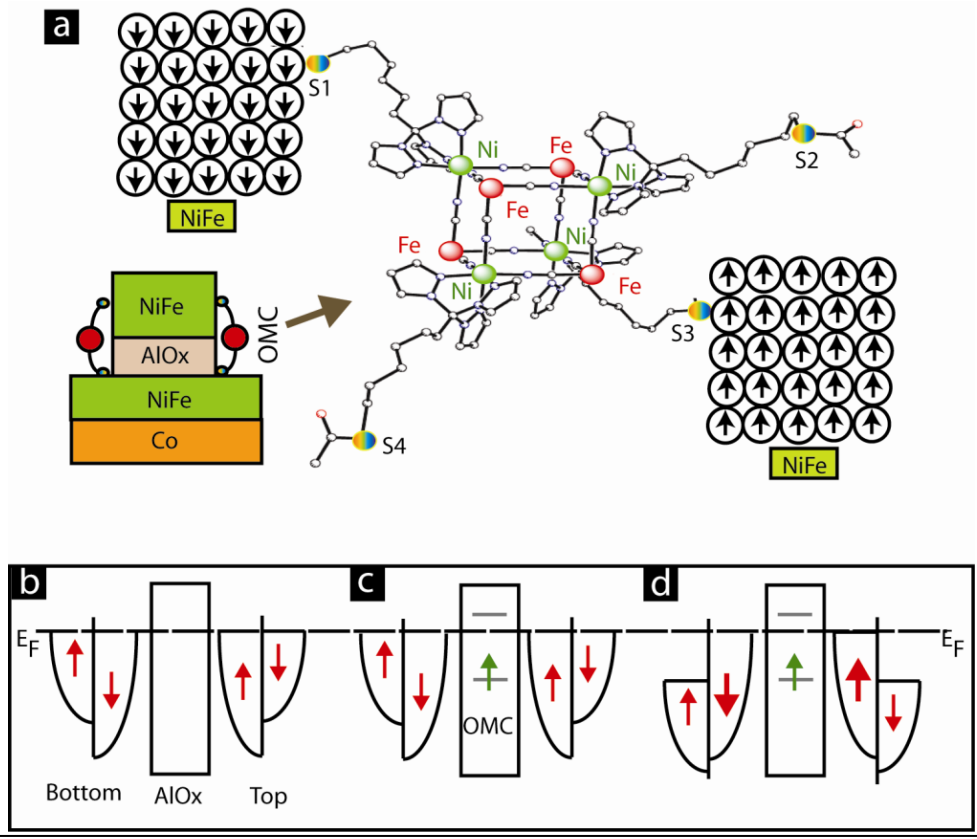

Fig. 22: Effect of OMC on spin bands: (a) a OMC bridge between the two FM electrodes, inset show the cross-sectional schematic of MEMSD. Spin band diagram of a MTJ (b) in bare state, (c) right after bridging of OMC channel, and (d) after the OMC induced rearrangement of spin bands to give highly spin polarized electrodes in the junction vicinity.

To understand the relation between the magnetic properties and the transport properties of the MEMSD one should explore the connection between *magnetic moment* and the P. In FM materials, P mainly depends on the DOS of itinerant sp electrons near $E_F$, whereas magnetic moment is dominated by the DOS of localized d electrons whose energy band reside below $E_F$ [40]. However, several alloys of Ni showed correlation between the P and magnetic moment [40]. Hence, it is highly likely that OMC can induce modification in the magnetic moment and transport, as both can be correlated with each other. This correlation is also expected to become strong when the two FM electrodes are strongly coupled to each other.

In the strong coupling limit, d electrons of FM electrodes presumably have much higher probability to move from one electrode to another via OMC channels (Fig. 22a). In the present case, an OMC channel possess a core containing eight d-orbitals atoms (four Ni and four Fe atoms) (Fig. 22a); importantly Ni atoms of an OMC are expected to connect with the Ni atoms of the FM electrodes (NiFe surface have Ni in elemental state and Fe in oxidized state [19]). Central cluster provide intermediate states to promote d- electron tunneling via alkane tethers (Fig. 22a). It is noteworthy that organic spin channels are far more superior to the metallic, insulator, and semiconducting materials [1]. Organic channels can conserve the spin for much longer distance and time. *In the case of MEMSD, strong d- electron coupling via OMC bridges are expected to make the FM electrodes' spatial wave function symmetric, as a result spin on the two FM electrodes become antisymmetric or antiparallel to each other.* In the present state, OMCs presumably work as a spin filter to populate single type of spin on each of the FM electrodes (Fig. 22a). We are unaware of the exact spin state of OMCs when connected to the two FM electrodes, but the current spin state is expected to make an OMC serving as a spin filter. We expect that a massive movement of spins via OMC channels occurred within the first few minutes or hours to produce highly spin polarized FM electrode region near junctions. Spin filtering may also be caused by the selective availability of tunneling paths through the OMCs, akin to single molecular magnets for which Berry phase [11, 48] and current suppression [49] were reported due to the selective spin path through them.

Effect of OMCs are expected to influence microscopic regions since FM electrodes possess long range magnetic ordering, helping spread the effect of OMCs induced strong exchange coupling in the junction area. Resultantly, molecular coupling affect the microscopic regions of FM electrodes in and around MTJ area (Fig. 20). Hence, OMCs were able to take over the transport through the planar AlOx tunnel barrier area by affecting FM electrodes. In simple terms, one can consider following analogy to understand the current suppression due to the changes in FM electrodes. Several orders change in a tunnel junction transport will occur if one can oxidize the metallic leads to become semiconducting. Oxidation of metallic leads is not affecting insulator of the tunnel junction, just affecting the density of states near Fermi level. In essence, OMCs channels made the magnetization of the two FM electrodes antiparallel, and presumably increased the P to make FM electrode 100% spin polarized. Under this state transport is prohibited through the planar area and the OMC bridges, producing current suppression phenomenon on MEMSDs.

## Summary

We have prepared molecular spin devices by attaching the magnetic molecular clusters on the exposed edges of a MTJ, to experimentally realize a MEMSD. In this study, 10,000 molecular bridges suppressed the current by 3-6 orders, below the leakage current level of the bare junction. Our magnetic studies evidenced that these dramatic change in MEMSD current is related to the equally staggering OMC induced changes in the properties of ferromagnetic electrodes. Magnetic studied on exactly the similar MEMSD configurations confirmed the dramatic effect of OMCs on the exchange coupling between ferromagnetic electrodes [33]. Evolution of extremely strong exchange coupling complements the dramatic changes in MEMSD current.

Other research groups have utilized different strategies to realize extremely strong exchange coupling between ferromagnetic films. For instance, exchange coupling was enhanced by reducing the thickness between two few monolayer thick ferromagnetic films to manipulate the Curie temperature like basic properties of the constituent films [23]. Efforts to manipulate exchange coupling can be done by inserting nanostructures in the nanogap between two ferromagnetic electrodes. Wong et al. [24] utilized iron nanoparticles to control the nature and magnitude of exchange coupling between two ferromagnetic films of MTJ, however, this study did not reveal a dramatic change in exchange coupling strength. However, molecular nanoparticles with net spin state produced rather dramatic observations. Pashupathy et al. [14] has experimentally showed that single or few $C_{60}$ molecules between Ni electrodes produced strong exchange coupling, leading to the observation of unprecedented Kondo level splitting at cryogenic temperature. In the case of MEMSDs thousands of molecules, covalently bonded on to the ferromagnetic electrodes, appear to amplify the molecule effect hence producing dramatic influence on the charge transport and exchange coupling at room temperature. Harnessing the strength of molecules as an exchange coupler between two ferromagnetic electrodes can lead to novel molecular spin devices.

MEMSD approach has a number of unique attributes to facilitate the fabrication of much vaunted molecular spin devices for numerous applications. MEMSD's salient features are following:

(i) *Molecule is a poor mechanical spacer and an excellent device element:* MEMSD has been mainly designed to utilize molecules as the device element. Nanometer separation between ferromagnetic electrodes is maintained by the alumina like insulator in a MTJ test bed. Previous studies utilizing molecules as the spacer and device elements were prone to short circuits and difficult to utilize magnetic electrodes [50]. Break junction approach utilize molecule as the device elements however, this approach requires sophisticated instrumentation and was mostly successful with gold [51, 52]. Producing breakjunction with different ferromagnetic material seems extremely challenging. Utilization of different magnetic electrodes is quite natural and effectively done in MTJs even at the commercial scale [40]. Transforming a MTJ into MEMSDs resolve most of the fabrication difficulties [50].

(ii) *Control experiments:* MEMSD allows a number of control experiments which are generally not possible with other approaches [50]. MTJs can be subjected charge/spin transport and magnetic characterization (Magnetization, FMR, and FMR etc.) before and after transforming it into molecular spin devices. Most importantly, one can undo the effect of molecule and retrieve the original device attributes. Molecules are bonded to MTJ test bed at the very end of device fabrication. Hence, molecule can interact with various chemicals and light radiation to provide deeper insights about device mechanism.

(iii) *Versatility:* Virtually any combination of magnetic electrodes can be connected to virtually any molecule, like single molecular magnets [5], porphyrin [6], DNA [53], nd organometallic molecular clusters [4]. Our MEMSDs were not able to switch between different current states in the controllable manner. To make MEMSD switchable between the two or more current states different combinations of MTJs and molecules should be attempted in future studies.

(iv) *New possibilities:* Molecular spin device paradigm is full of already calculated [3, 10, 11, 14, 54] and undiscovered phenomenon. For instance, single molecular magnets are considered highly promising for realizing quantum computational devices [3, 5]. MEMSD approach can host such molecules as a device element for exploring the potential of realizing next generation computational devices.

(v) *Improvement in MEMSD fabrication approach:* The liftoff based approach used for preparing exposed side edges for molecule attachment is simple but generally gave low yield. Low yield was caused by the higher thickness of insulator or spikes along the edges, which depended on the quality of photoresist and photolithography step. Molecules are unable to bridge if effective insulator thickness is more than the physical length. A brief chemical etch to remove 50-100 nm insulator width may greatly enhance the possibility of transforming MTJs into MEMSDs. Etching insulator without affecting metal electrodes is regularly accomplished in the synthesis of nanowire [55]. There are several other approaches to produce multilayer edge molecular devices, similar to the one discussed in this paper [50]; however, none of them employed ferromagnetic electrodes to realize molecular spin devices.

Further research on MEMSD systems by independent research group will play crucial role in establishing the results discussed in this paper. During the research on MEMSDs large variability in experimental conditions made it extremely challenging to perform complementary studies for longer duration. Extensive efforts were made to produce stable and molecular solution resistant MTJ test beds [18, 20]. However, realizing the possibility of unnoticed artifact, *especially in the charge transport measurements*, it is recommended that any potential euphoria about this paper should be restrained until other independent researchers confirm or support these findings.

**Acknowledgments:**

PT thanks Prof. Bruce J. Hinds and the Department of Chemical and Materials Engineering, University of Kentucky to enable his PhD research work presented in this manuscript. He also thanks Prof. D.F Li and Prof. S. M. Holmes for providing molecules used in this work. Discussion in this paper solely reflects PT's views and perspective. PT dedicates this paper to the late Mr. George Spiggle, a staff engineer at University of Kentucky, for maintaining the research facilities and promptness in fixing numerous equipment problems.